\documentclass[twoside,twocolumn,9pt]{article}
\usepackage{caption}
\usepackage{subcaption}
\usepackage{extsizes}
\usepackage[super,sort&compress,comma]{natbib} 
\usepackage[version=3]{mhchem}
\usepackage[left=1.5cm, right=1.5cm, top=1.785cm, bottom=2.0cm]{geometry}
\usepackage{balance}
\usepackage{times,mathptmx}
\usepackage{sectsty}
\usepackage{graphicx} 
\usepackage{lastpage}
\usepackage[format=plain,justification=justified,singlelinecheck=false,font={stretch=1.125,small,sf},labelfont=bf,labelsep=space]{caption}
\usepackage{float}
\usepackage{fancyhdr}
\usepackage{fnpos}
\usepackage[english]{babel}
\addto{\captionsenglish}{%
  
}
\usepackage{array}
\usepackage{droidsans}
\usepackage{charter}
\usepackage[T1]{fontenc}
\usepackage[usenames,dvipsnames]{xcolor}
\usepackage{setspace}
\usepackage[compact]{titlesec}
\usepackage{hyperref}
\usepackage{braket}
\def\ben{\begin{equation}}
\def\een{\end{equation}}

\def\LP{\rm LP}

\def\UP{\rm UP}
\def\D{\rm D}

\def\hc{\text{h.c.}}
\def\B{\text{B}}

\def\nonum{\nonumber \\}

\usepackage{epstopdf}

\definecolor{cream}{RGB}{222,217,201}

\begin{document}

\pagestyle{fancy}
\thispagestyle{plain}
\fancypagestyle{plain}{

\fancyhead[C]{\includegraphics[width=18.5cm]{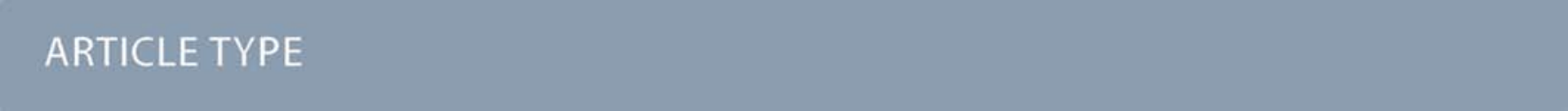}}
\fancyhead[L]{\hspace{0cm}\vspace{1.5cm}\includegraphics[height=30pt]{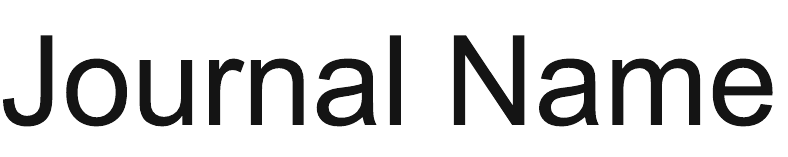}}
\fancyhead[R]{\hspace{0cm}\vspace{1.7cm}\includegraphics[height=55pt]{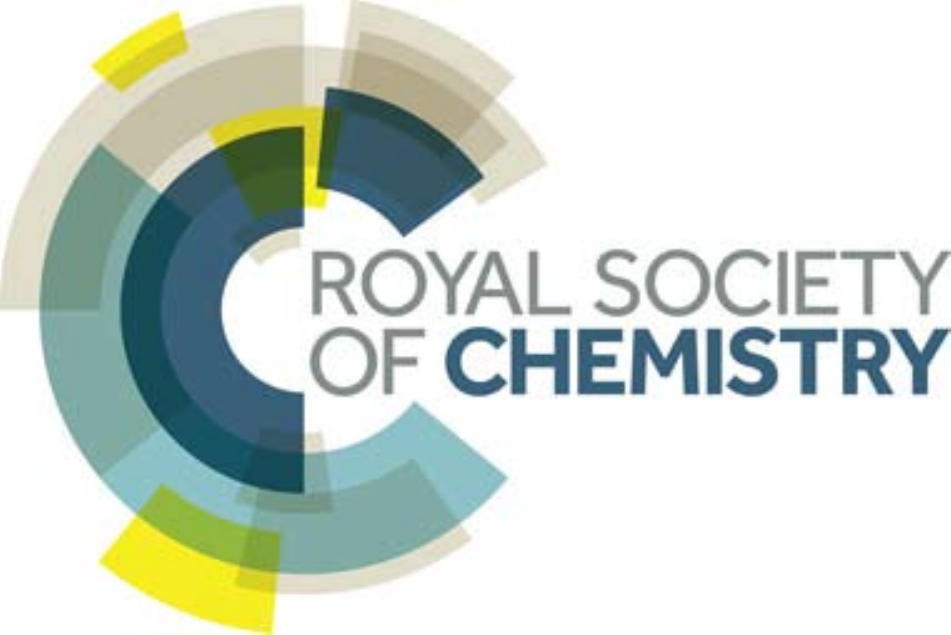}}
\renewcommand{\headrulewidth}{0pt}
}

\makeFNbottom
\makeatletter
\renewcommand\LARGE{\@setfontsize\LARGE{15pt}{17}}
\renewcommand\Large{\@setfontsize\Large{12pt}{14}}
\renewcommand\large{\@setfontsize\large{10pt}{12}}
\renewcommand\footnotesize{\@setfontsize\footnotesize{7pt}{10}}
\makeatother

\renewcommand{\thefootnote}{\fnsymbol{footnote}}
\renewcommand\footnoterule{\vspace*{1pt}%
\color{cream}\hrule width 3.5in height 0.4pt \color{black}\vspace*{5pt}} 
\setcounter{secnumdepth}{5}

\makeatletter 
\renewcommand\@biblabel[1]{#1}            
\renewcommand\@makefntext[1]%
{\noindent\makebox[0pt][r]{\@thefnmark\,}#1}
\makeatother 
\renewcommand{\figurename}{\small{Fig.}~}
\sectionfont{\sffamily\Large}
\subsectionfont{\normalsize}
\subsubsectionfont{\bf}
\setstretch{1.125} 
\setlength{\skip\footins}{0.8cm}
\setlength{\footnotesep}{0.25cm}
\setlength{\jot}{10pt}
\titlespacing*{\section}{0pt}{4pt}{4pt}
\titlespacing*{\subsection}{0pt}{15pt}{1pt}

\fancyfoot{}
\fancyfoot[LO,RE]{\vspace{-7.1pt}\includegraphics[height=9pt]{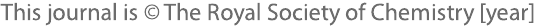}}
\fancyfoot[CO]{\vspace{-7.1pt}\hspace{13.2cm}\includegraphics{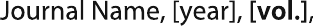}}
\fancyfoot[CE]{\vspace{-7.2pt}\hspace{-14.2cm}\includegraphics{RF}}
\fancyfoot[RO]{\footnotesize{\sffamily{1--\pageref{LastPage} ~\textbar  \hspace{2pt}\thepage}}}
\fancyfoot[LE]{\footnotesize{\sffamily{\thepage~\textbar\hspace{3.45cm} 1--\pageref{LastPage}}}}
\fancyhead{}
\renewcommand{\headrulewidth}{0pt} 
\renewcommand{\footrulewidth}{0pt}
\setlength{\arrayrulewidth}{1pt}
\setlength{\columnsep}{6.5mm}
\setlength\bibsep{1pt}

\makeatletter 
\newlength{\figrulesep} 
\setlength{\figrulesep}{0.5\textfloatsep} 

\newcommand{\topfigrule}{\vspace*{-1pt}%
\noindent{\color{cream}\rule[-\figrulesep]{\columnwidth}{1.5pt}} }

\newcommand{\botfigrule}{\vspace*{-2pt}%
\noindent{\color{cream}\rule[\figrulesep]{\columnwidth}{1.5pt}} }

\newcommand{\dblfigrule}{\vspace*{-1pt}%
\noindent{\color{cream}\rule[-\figrulesep]{\textwidth}{1.5pt}} }

\makeatother

\twocolumn[
  \begin{@twocolumnfalse}
\vspace{3cm}
\sffamily
\begin{tabular}{m{4.5cm} p{13.5cm} }

\includegraphics{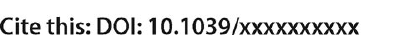} & \noindent\LARGE{\textbf{Polariton Chemistry: controlling molecular dynamics with optical cavities}} \\

\vspace{0.3cm} & \vspace{0.3cm} \\

 & \noindent\large{Raphael F. Ribeiro\textit{$^{a}$}, Luis A. Mart\'inez-Mart\'inez\textit{$^{a}$}, Matthew Du\textit{$^{a}$}, Jorge Campos-Gonzalez-Angulo\textit{$^{a}$}, Joel Yuen-Zhou$^{\ast}$\textit{$^{a}$}} \\
 
\includegraphics{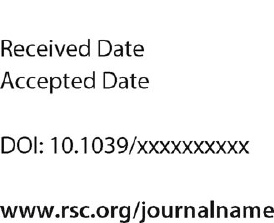} & \noindent\normalsize{Molecular polaritons are the optical excitations which emerge when molecular transitions interact strongly with confined electromagnetic fields. Increasing interest in the hybrid molecular-photonic materials that host these excitations stems from recent observations of their novel and tunable chemistry. Some of the remarkable functionalities exhibited by polaritons include the ability to induce long-range excitation energy transfer, enhance charge conductivity, and inhibit or enhance chemical reactions. In this review, we explain the effective theories of molecular polaritons which form a basis for the interpretation and guidance of experiments at the strong coupling limit. The theoretical discussion is illustrated with the analysis of innovative applications of strongly coupled molecular-photonic systems to chemical phenomena of fundamental importance to future technologies.} \\
\end{tabular}

 \end{@twocolumnfalse} \vspace{0.6cm}

  ]

\renewcommand*\rmdefault{bch}\normalfont\upshape
\rmfamily
\section*{}
\vspace{-1cm}

\footnotetext{\textit{$^{a}$~Department of Chemistry and Biochemistry, University of California San Diego, La Jolla, California 92093, United States. E-mail: jyuenzhou@ucsd.edu}}



\section{Introduction}

Light plays a fundamental role in chemistry. It is an essential ingredient in many biological and synthetic chemical reactions and a basic tool for the investigation of molecular properties \cite{turro1991modern, balzani2014photochemistry}. However, in most cases where photons participate in a chemical event, the interaction of light and matter is weak enough that it may be treated as a small perturbation. In this \textit{weak-coupling} regime, radiation provides only a gateway for a molecular system to change its quantum state. This paradigm is the basis of spectroscopy and photochemical dynamics, but fails entirely when the interaction of a single photon with many molecules is intense enough to overcome their dissipative processes. In this case, just as strongly-interacting atoms form molecules, hybrid states (\textit{molecular polaritons}) with mixed molecular and photonic character result from the strong coupling of the electromagnetic (EM) field with molecules \cite{hopfield_theory_1958, agranovich1960dispersion, ebbesen_hybrid_2016}. In this regime, both photons and molecular excited-states lose their individuality, just as atoms do when they form molecules. 

The generic setup of a device with polaritonic excitations consists of a medium that confines electromagnetic fields to the microscale (hereafter we will refer to this as an optical \textit{microcavity} \cite{kavokin2017microcavities, baranov2018}, although strong coupling has also been achieved with metal layers supporting plasmons \cite{torma_strong_2015, vasa2018}) and a condensed-phase molecular ensemble with one or more \textit{bright} (optical) transitions nearly-resonant with the optical cavity. Typically, molecular transitions can be well approximated to have no wave vector dependence (since optical wavelengths are much larger than molecular length scales) with linewidth dominated by the coupling to intra and intermolecular degrees of freedom. Conversely, the microcavity spectra shows a well-defined wave vector dependence (Fig. \ref{cavmod}) and homogeneous broadening due to the weak interaction with the external EM fields \cite{kavokin2017microcavities}. The strong coupling regime is achieved when the rate of energy exchange between microcavity photons and the molecular system is larger than the rate of cavity-photon leakage and molecular dephasing. In this case, the elementary optical excitations (quasiparticles) of the heterostructure consist of superpositions of delocalized (coherent) molecular states and photonic degrees of freedom, with energies and lifetimes which can significantly differ from those of the bare excitations. The capacity to tune the energies and photon/molecular content of polariton states is a main attraction of the strong coupling regime  (Sec. \ref{bas}). Nevertheless, the hybrid cavity also contains a large number of incoherent, or \textit{dark} molecular states, which may be more or less localized depending on disorder \cite{houdre_vacuum-field_1996, agranovich2003}, and geometry \cite{gonzalez-ballestero2016} of the hybrid material. Despite their weak contribution to the optical response, the dark states are fundamentally important for a description of the novel chemical dynamics emergent at the strong coupling regime.

While theoretical studies of hybrid states of light and matter date back to the 1950s \cite{hopfield_theory_1958, agranovich1960dispersion}, and observations of atomic and solid-state cavity-polaritons first happened in the 1980s \cite{meschede1985, raizen1989} and 1990s \cite{weisbuch_observation_1992, houdre_room_1993}, respectively,  it is only recently that experimental \cite{lidzey1998, schouwink2001, holmes2004a, dintinger_strong_2005, kena-cohen_strong_2008, kena-cohen2010, virgili2011, schwartz2011, aberra_guebrou_coherent_2012, hutchison_modifying_2012, schwartz_polariton_2013, simpkins_spanning_2015, thomas_ground-state_2016, zhong2017, chikkaraddy2016, melnikau2017, baieva_dynamics_2017, crum2018, rozenman2018, dunkelberger2018b, cheng2018} and theoretical \cite{cwik2014, pino2015, herrera2016, galego2016, kowalewski2016, bennett2016a, wu2016, herrera_dark_2017, zhang2017, feist2017, dimitrov2017, flick_cavity_2017, zeb2018, yuen-zhou2017, martinez-martinez2018, du2017, martinez-martinez2017b, ribeiro2017c} activity have flourished in the field of strongly coupled chemistry. This attention can be attributed in part to the experimental observations of polariton effects on chemical dynamics, which thus offer novel pathways for the control of molecular processes \cite{ebbesen_hybrid_2016}. 
 
In this review we provide a theoretical perspective on the recent advances in molecular polaritons arising from electronic (\textit{organic exciton-polaritons}) or vibrational (\textit{vibrational-polaritons}) degrees of freedom. Our discussions are primarily based on quantum-mechanical effective models, which are general enough to be applied in regimes where a classical description is inaccurate \cite{carusotto_quantum_2013}. Furthermore, the models discussed here describe only the relevant low-energy degrees of freedom probed by experiments. This allows a consistent and predictive description of the behavior of strongly coupled ensembles including a macroscopic number of molecules. First-principles approaches are explored in Refs. \cite{ruggenthaler2014, dimitrov2017, flick_cavity_2017}. Finally, it is not our intent to provide a complete review of the fast-growing molecular polariton literature. We have decided to present the basic theory and illustrate it with examples, that we believe, show general principles that might be useful for future investigations of polariton chemistry. For reviews on other aspects of molecular polaritons not emphasized here, see Refs. \cite{holmes2007, agranovich2011, kena-cohen2012, torma_strong_2015, ebbesen_hybrid_2016, sukharev_optics_2017, herrera2018, vasa2018, s.dovzhenko2018}. 

This review is organized as follows. In Sec. \ref{bas} we present the general concepts that form the basis for molecular polaritonics. Sec. \ref{orgp} provides an overview of the theory of organic exciton-polaritons. This is illustrated with applications to polariton-mediated chemical reactivity (Sec. \ref{orgappI}), energy transfer (Sec. \ref{orgappII}) and singlet fission (Sec. \ref{orgappIII}). In Sec. \ref{vibp} we discuss the theory and phenomenology of vibrational-polaritons. We focus on the effects of vibrational anharmonicity on their nonlinear response, and revisit exciting experimental results probing the thermodynamics and kinetics under vibrational strong coupling in Secs. \ref{vib_trans} and \ref{vib_app}, respectively. The ultrastrong regime of light-matter interaction is briefly introduced in Sec. \ref{ultra}. This review is concluded in Sec. \ref{epi}. 

\section{Polariton basics}\label{bas}

We introduce the basic notions of polariton behavior in this Sec. by examining the simplest models displaying strong coupling between light and matter. The bare microcavity modes are reviewed in Sec. \ref{optspec}, and the spectrum resulting from the strong coupling of two-level systems with a single cavity mode is discussed in Sec. \ref{jctcs}. The intuition given by the results discussed in this section will guide all later developments.

\subsection{Optical microcavity spectra}\label{optspec}
\begin{figure}
\includegraphics[width=\columnwidth]{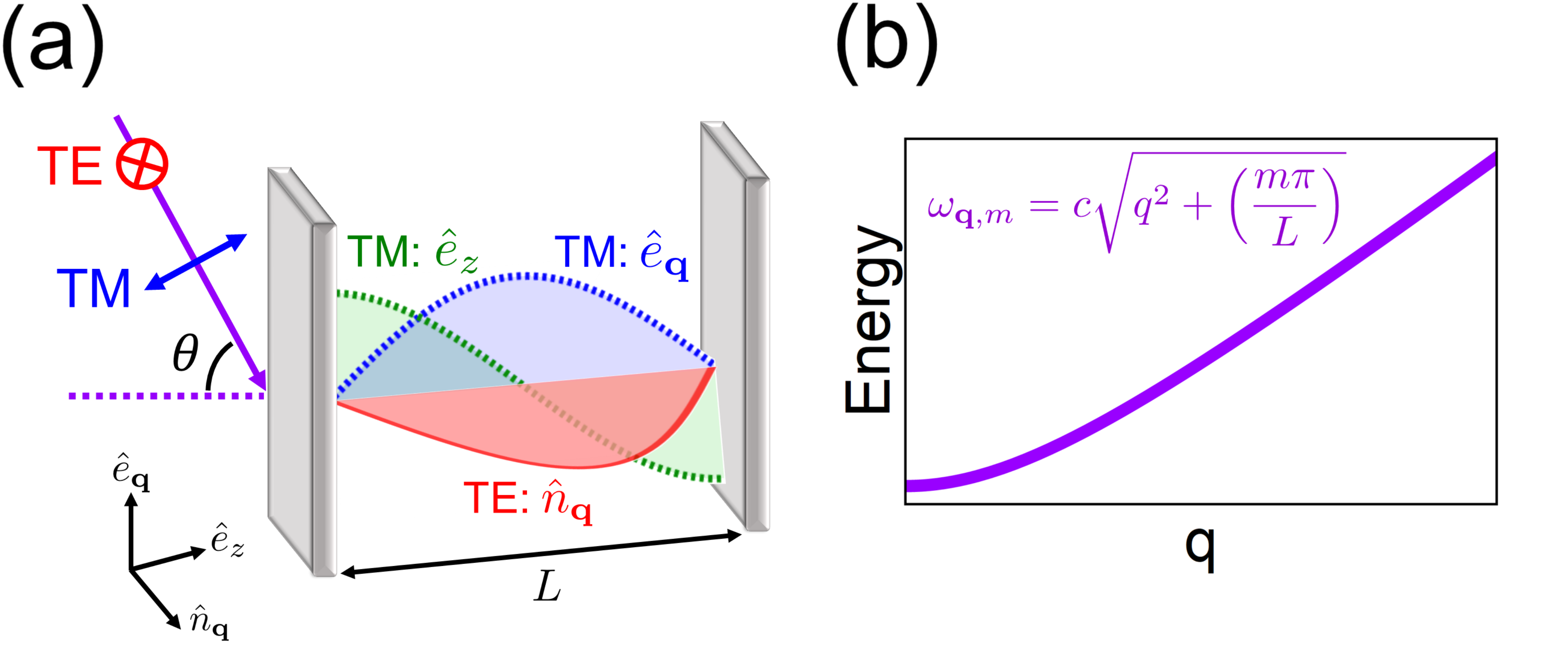}
\caption{(a) Representation of microcavity modes excited by radiation (purple)
incident at angle $\theta$. The electric field polarization of the TE 
modes (red) lies along $\hat{n}_{\mathbf{q}}$ while that of TM (blue and green) has $\hat{e}_{\mathbf{q}}$ and  $\hat{e}_{z}$ components. (b)
Dispersion (energy as a function of mirror-plane wave vector $\mathbf{q}$)
of the photonic mode of index $m$ in microcavity with transverse length $L$. }\label{cavmod}
\end{figure}

The optical cavities employed for molecular strong coupling studies generally consist of two highly-reflective (at the frequencies of interest) parallel metallic or dielectric mirrors separated by a distance $L$ on the order of $\mu$m \cite{kavokin2017microcavities}. The length of the cavity is typically chosen to be resonant with a molecular transition. The EM modes of these devices are classified by the in-plane wave vector $\mathbf{q}$, the integer band number $m$ (where $q_z = m\pi/L$) associated with the transverse confinement direction, and the electric field polarization [transverse magnetic (TM) or transverse electric (TE)] \cite{steck2007quantum, kavokin2017microcavities} (see Fig.\ref{cavmod}a).  The TE polarization is perpendicular to the incidence plane, while the TM  belongs to it. The former vanishes at the mirrors, in contrast to the latter. However, in lossless microcavities, the TM and TE modes are degenerate, and when $|\mathbf{q}| \rightarrow 0$ their spatial distributions become identical. The microcavity is typically engineered to have a single band (Fig. \ref{cavmod}b) containing a resonance with the material (though there exist exceptions \cite{simpkins_spanning_2015, coles2014}). The remaining bands are highly off-resonant, and thus, may be neglected in a low-energy theory of polaritons. It is also sometimes useful to perform a long wavelength approximation which disregards the spatial variation of the electric field, and includes explicitly only a single microcavity mode which interacts with a macroscopic collection of optically-active molecular transitions. This is appropriate whenever the density of accessible molecular states is much larger than the photonic (total internal reflection of incident radiation with $\theta > \theta_c$ (Fig. \ref{cavmod}a) establishes a natural cut-off frequency for the cavity modes which can be accessed by excitation with external radiation; alternatively, a cutoff can be imposed on cavity photons which are highly-detuned from the molecular transition \cite{pino2015, Daskalakis2017}). Such condition is fulfilled in most observations of molecular polaritons in condensed-phase media.

\subsection{Jaynes-Cummings and Tavis-Cummings models}\label{jctcs}
\begin{figure}
\includegraphics[width=\columnwidth]{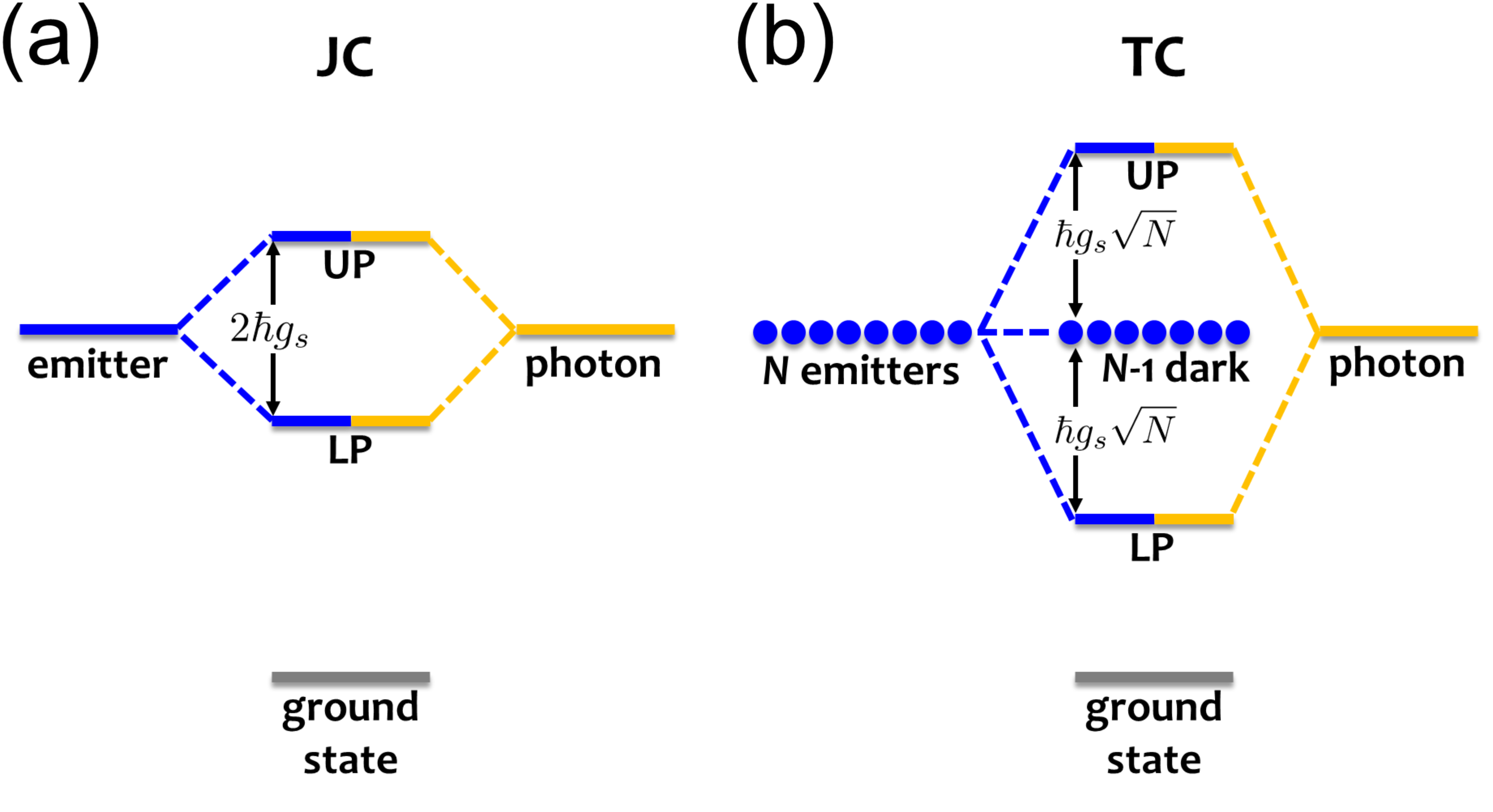}
\caption{(a) Jaynes-Cummings (JC) model: emitter and photon strongly couple
to form hybridized states termed lower and upper polaritons (LP and
UP, respectively) separated in energy by $2\hbar g_{s}$;
(b) Tavis-Cummings (TC) model: $N$ emitters interact strongly with a photon to yield 
polariton (LP, UP) and $N-1$ dark states. The latter do not couple to light and thus maintain the original emitter energy.}\label{jctc}\end{figure}

When we introduce a bright two-level system to a single-mode microcavity we obtain the Jaynes-Cummings (JC) model \cite{jaynes_comparison_1963}. In particular, it consists of a \textit{lossless} cavity-mode of frequency $\omega_c$ interacting with a two-level system of transition energy $\hbar\omega_s$. Thus, the effective Hamiltonian of the JC model is given by

\begin{equation} H_{\text{JC}}  =  \hbar \omega_c a^\dagger a + \hbar \omega_s \sigma_+ \sigma_- - \hbar g_s\left(a^\dagger \sigma_- + a\sigma_+\right), \label{jceq}\end{equation} 
where $a \left(a^\dagger\right)$ is the cavity photon annihilation (creation) operator, $\sigma_+$ $(\sigma_-)$ creates  (annihilates) excitons, and $\hbar g_s = \mu \cdot E\sqrt{\hbar \omega_c/2\epsilon V_c}$ is the strength of the radiation-matter interaction, where $\epsilon$ is the dielectric constant of the intracavity medium, $V_c$ is the effective mode volume of the (cavity) photon \cite{ujihara1991}, $E$ is the photon electric field amplitude at the emitter position, and $\mu$ is the transition dipole moment of the latter. Notably, this Hamiltonian implicitly assumes that the light-matter interaction is \textit{strong} relative to the damping of each degree of freedom, yet \text{weak} compared to both $\omega_c$ and $\omega_s$. Thus, only states with equal total number of excitations \ben N_{\text{exc}} = a^\dagger a + \sigma_+ \sigma_- \een
are coupled by the light-matter interaction. As a result, the ground-state of the hybrid system is equivalent to that of the decoupled. The same is clearly not true for the excited-states. For any $N_{\text{exc}} = N > 0$,  $H_{\text{JC}}$ has two hybrid photon-matter eigenstates. For example, the lowest-lying excited-states of the system have $N_{\text{exc}} = 1$. They are given by (Fig. \ref{jctc}a)
\begin{align} & \ket{\LP} = \text{cos}(\theta_{\text{JC}})\ket{g,1}+\text{sin}(\theta_{\text{JC}})\ket{e,0}, \nonum & \ket{\UP} = -\text{sin}(\theta_{\text{JC}})\ket{g,1} + \text{cos}(\theta_{\text{JC}})\ket{e,0}, \label{jcp}\end{align}
where $\ket{g,N} (\ket{e,N})$ denotes a state where the material is in the ground(excited)-state and the cavity has $N$ photons, and $\theta_{\text{JC}} = \text{tan}^{-1}\left[2g_s/(\omega_c-\omega_s)\right]$ is the polariton mixing-angle, which determines the probability amplitude for a photon or emitter to be observed when the state of the system is either $\ket{\text{LP}}$ or $\ket{\text{UP}}$. The state $\ket{\LP}$ is called \textit{lower polariton}, while $\ket{\UP}$ is the \textit{upper polariton}. Their energy difference is $\hbar\Omega_R = 2\hbar\sqrt{\Delta^2/4+g_s^2}$, where $\Delta = \omega_c - \omega_s$ is the \textit{detuning} between the photon and emitter frequencies. At resonance $(\omega_c = \omega_s)$, the LP and UP become a maximally entangled superposition of emitter and cavity photon, with \textit{vacuum Rabi splitting} $\hbar\Omega_R = 2\hbar g_s$ [the terminology refers to the process by which introduction of an emitter to the cavity \textit{vacuum} (initial-state $\ket{e,0}$) leads to coherent (Rabi) oscillations with frequency $\Omega_R/2$ in the probability to detect a photon or a material excited-state inside the cavity \cite{rabi1937, agarwal1984}]. For positive detunings, the UP (LP) has a higher photon (emitter) character, while the opposite is true when $\omega_c < \omega_s$. The JC model shows other interesting features, such as photon blockade \cite{birnbaum2005}. However, because it contains no more than a \textit{single} structureless [\textit{i.e.}., the internal (vibronic, vibro-rotational, etc) structure of molecular excitations is not considered] emitter, the JC model is only of pedagogical significance for chemistry, although conditions in which a single-molecule is strongly coupled to a microcavity have only been achieved in few studies \cite{chikkaraddy2016, benz_single-molecule_2016, wang2017}. In fact, for reasons we will discuss next, most experiments which probe the strong coupling regime employ a molecular ensemble including a macroscopic number of emitters.

The generalization of the JC model for the case where $N$ \textit{identical} two-level emitters interact strongly with a lossless cavity mode is denoted the Tavis-Cummings (TC) \cite{tavis_exact_1968, tavis_approximate_1969} or Dicke model \cite{dicke_coherence_1954}.
It is described with the Hamiltonian
\begin{equation} H_{\text{TC}}  =  \hbar \omega_c a^\dagger a + \hbar \omega_s\sum_{i=1}^N \sigma_+^{(i)}\sigma_-^{(i)} - \hbar g_s \sum_{i=1}^N \left(a^\dagger \sigma_-^{(i)} + a\sigma_+^{(i)}\right), \label{tceq}\end{equation}  
where the superscript $i$ labels each of the $N$ emitters. Note that while $N$ can be very large, the emitters are assumed to occupy a region of space where the variation of the electric field amplitude can be neglected.  The spectrum of $H_{\text{TC}}$ differs markedly from that of $H_{\text{JC}}$. However, the total number of excitations of the system $N_{\text{exc}} = a^\dagger a  + \sum_{i=1}^N \sigma_+^{(i)}\sigma_-^{(i)}$ remains a constant of motion. Thus, similar to $H_{\text{JC}}$,  the TC model only allows hybrid states in which all components share the same $N_{\text{exc}}$. Specifically, there exists $N+1$ basis states with $N_{\text{exc}} = 1$: a single state with all emitters in the ground-state and a cavity photon, $\ket{g,1} \equiv \ket{0,0,0...,0;1} $, and $N$ states where a single-molecule is excited and the cavity EM field is in its ground-state, $\ket{e^{(i)},0} \equiv \ket{0,0,...,e^{(i)},...,0,0;0}, i \in \{1,...,N\}$. The stationary states with $N_{\text{exc}}=1$ (Fig. \ref{jctc}b) not only include the polaritons, but also a degenerate manifold of \textit{dark states} $\ket{\D_\mu},~\mu \in \{1,...,N-1\}$, for which an orthogonal basis may be given by delocalized \textit{non-totally-symmetric} (under any permutation of the emitters) molecular excited-states orthogonal to the \textit{permutationally-invariant} bright state \cite{vetter2016single, agarwal1998}. This can be easily seen by rewriting $H_{\text{TC}}$ in terms of bright- and dark state operators, $\sigma^{(\B)} = \frac{1}{\sqrt{N}}\sum_{i=1}^{N} \sigma^{(i)}$ and $\sigma^{(\D_\mu)}$, respectively, where $\sigma$ can be any operator acting on the total Hilbert space of the two-level system ensemble, and the normalization is chosen so that the commutation relations of the $\sigma$ matrices are preserved. In this basis, $H_{\text{TC}}$ is given by
\begin{equation} H_{\text{TC}}  =  \hbar \omega_c a^\dagger a + \hbar \omega_s \sigma_+^{(\B)}\sigma_-^{(\B)} - \hbar \sqrt{N}g_s \left(a^\dagger \sigma_-^{(\B)} + a\sigma_+^{(\B)}\right) + H_{\D}, \label{htc} \end{equation}  
where $H_{\D} = \hbar \omega_s  \sum_{\mu=1}^{N-1} \sigma_+^{(\D_\mu)}\sigma_-^{(\D_\mu)}$ is the dark Hamiltonian. From Eq. \ref{htc} and its similarity with Eq. \ref{jceq}, it is clear that the hybrid eigenstates with $N_{\text{exc}}=1$ are given by
\begin{align} & \ket{\LP} = \text{cos}(\theta_{\text{TC}})\ket{g,1}+\text{sin}(\theta_{\text{TC}})\ket{\B,0}, \nonum
 & \ket{\UP} = -\text{sin}(\theta_{\text{TC}})\ket{g,1} + \text{cos}(\theta_{\text{TC}})\ket{\B,0}, \end{align}
 where $\ket{\B,0}$ is the totally-symmetric bright emitter state \ben \ket{\B,0} = N^{-1/2}\sum_{i=1}^N\ket{e^{(i)},0}, \een and $\theta_{\text{TC}} = \text{tan}^{-1}\left[2\sqrt{N}g_s/(\omega_c-\omega_s)\right]$. Notably, these states are simple generalizations of the JC polaritons provided in Eq. \ref{jcp}. However, the vacuum Rabi splitting given by Eq. \ref{htc} is significantly enhanced compared to JC, as a result of the \textit{collective} light-matter coupling $g_s \sqrt{N}$ which couples a cavity photon to a delocalized bright emitter. In fact, at resonance, the (collective) vacuum Rabi splitting in the TC model is given by $\hbar\Omega_R = 2\sqrt{N}g_s$. Since $g_s \propto V_c^{-1/2}$, it follows that $\hbar\Omega_R$ scales with the density of emitters in the optical mode volume. Thus, it is much easier to reach strong coupling between light and matter with a large concentration of optically-active material.

\begin{figure}
\includegraphics[width=\columnwidth]{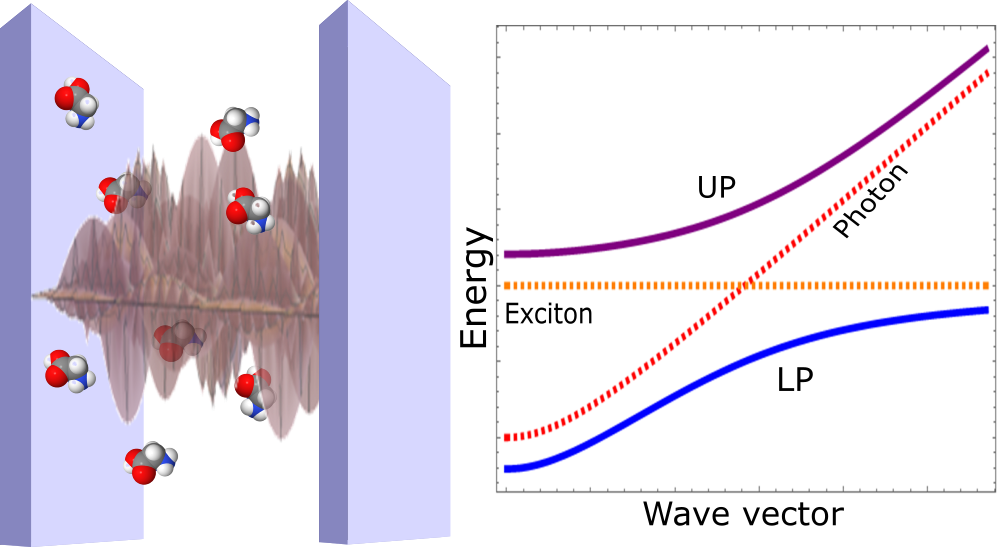}
\caption{The l.h.s. gives a pictorial representation of a set of molecular emitters embedded in a resonant planar microcavity; the r.h.s presents the cavity, emitter and polariton dispersions $[\text{energy in terms of wave vector} (\mathbf{q})]$ according to the multimode generalization of the TC model.}\label{orgsk}\end{figure}

Introduction of emitter disorder and cavity losses to the JC and TC models does not change the essential conclusions of the above discussion as long as $\hbar\Omega_R$ remains larger than the broadening due to the photonic and emitter damping. For instance, while inhomogeneous broadening breaks the degeneracy and delocalization of the dark manifold and leads to photonic transfer (intensity borrowing) from the LP and UP to these states, the fraction of transferred photon is proportional to $|\eta/\hbar \Omega_R|^2$, where $\eta$ is the energetic width of the inhomogeneous disorder. For $\hbar \Omega_R \gg \eta$, the photonic contribution to the TC dark states is very small and may be neglected in most cases \cite{houdre_vacuum-field_1996} (though it can be significant in cavity-absorption measurements). The same analysis allows us to conclude that inhomogeneous broadening is suppressed in polariton spectra. In fact, this has been repeatedly observed in atomic and inorganic semiconductor quantum well cavity-polaritons where the emitters are approximately structureless (see e.g., \cite{manceau2017}). 

We emphasize the above discussion disregards any dependence of the hybrid cavity Hamiltonian on real-space position or wave vector. In reality, the emitters are spatially distributed within the cavity volume and interact differently with the cavity-mode continuum (Fig. \ref{cavmod}) according to their positions (in the case of molecular systems there is also a dependence on the orientation of the transition dipole moment). Thus, the LP and UP define bands with dispersion given by $\omega_{\LP}(\mathbf{q})$ and $\omega_{\UP}(\mathbf{q})$, respectively (Fig. \ref{orgsk}). The effects of disorder are more complex in this case, since they may lead to strong polariton localization and scattering \cite{agranovich2003, litinskaya_loss_2006, agranovich_nature_2007, litinskaya_propagation_2008}. Nevertheless, if the emitter density of states (DOS) is much larger than that of the cavity EM field, then the  dark molecular modes still constitute the majority of the states of the hybrid microcavity. This observation is crucial for the investigation of relaxation dynamics in molecular polaritons, and we devote more attention to it in Sec. \ref{orgrel}.

The aspects of the strong coupling regime discussed in this Sec. are essential for a description of molecular polaritons. Nonetheless, the TC model is still too primitive for most chemistry purposes since the two-level systems mimicking electronic states carry no internal (for example, vibronic) structure which is fundamental for the description of molecular dynamics.

\section{Organic exciton-polaritons}\label{orgp}
Organic semiconductors are suitable materials for strong coupling due to their large transition dipole moments and sufficiently narrow linewidths \cite{agranovich2009excitations}. In fact, the first observations of molecular cavity-polaritons originated from the coupling of an optical microcavity with the organic excitons of a Zn-porphyrin dye \cite{lidzey1998}, and later with J-aggregate films \cite{lidzey1999}. Recent years have seen many remarkable developments including demonstrations of reversible optical switching \cite{schwartz2011}, suppression of photochemical reactivity \cite{hutchison_modifying_2012}, room-temperature polariton lasing \cite{kena-cohen2010} and Bose-Einstein condensation \cite{plumhof_room-temperature_2014},  enhanced charge conductivity \cite{orgiu_conductivity_2015}, and long-range excitation energy transfer \cite{coles2014b, zhong2017, georgiou2018}. We introduce the effective Hamiltonian of organic polaritons in Sec. \ref{orgbas}, review their relaxation dynamics in Sec. \ref{orgrel}, and discuss some of their applications in light of the presented theoretical framework in Secs \ref{orgappI}, \ref{orgappII}, and \ref{orgappIII}.

\begin{table*}
\caption{Timescales relevant for the description of organic (J-aggregate) microcavity relaxation dynamics}
\label{tbl:compare}\centering %
\begin{tabular}{>{\centering}m{3.5cm}>{\centering}m{3cm}>{\centering}m{3cm}>{\centering}m{3.0cm}>{\centering}m{2.0cm}}
\hline 
Process & Initial state(s) & Final state(s) & Timescale & Ref.\tabularnewline
\hline 
Rabi splitting & \textemdash{} & \textemdash{} & $15-50~\rm{fs}$
($80-300~\rm{meV}$) & \cite{hobson2002}\tabularnewline
\hline
Cavity leakage & cavity photon & \textemdash{} & $ 35-100~\rm{fs}$ & \cite{coles2011a, michetti2008}\tabularnewline \hline
 & exciton & \textemdash{} & $10-1000~\rm{fs}$ & \cite{taylor1984, reiser1982} \tabularnewline 
 & UP & incoherent excitons & $\sim 50~\rm{fs}$ & 
\cite{agranovich2003}\tabularnewline 
 Vibrational relaxation & incoherent excitons & LP & $\sim 10~\rm{ps}$ & \cite{litinskaya2004}\tabularnewline
 & incoherent excitons & UP & $\sim 1~\mu$s & \cite{coles2011}\tabularnewline
 \hline
& UP & \textemdash{} & $\sim 100$ fs & \cite{michetti2008}\tabularnewline
 Photoluminescence & LP & \textemdash{} & $\sim 1 ~\rm{ps}$ & \cite{wang2014}\tabularnewline
 & bare exciton & \textemdash{} & $\sim 1-20 ~\rm{ps}$ & \cite{furuki2001, song2004, schwartz_polariton_2013, wang2014}\tabularnewline \hline  
\end{tabular}
\end{table*}

\subsection{Effective descriptions} \label{orgbas}
The main novelty introduced by organic (Frenkel) exciton-polaritons \cite{may2004charge, agranovich2009excitations} is the significant local \textit{vibronic coupling} of the molecular excited-states with inter and intramolecular vibrational modes. This gives rise to inhomogeneously broadened linewidths, vibronic progressions, and Stokes shifts in the optical spectra of organic systems \cite{may2004charge, agranovich2009excitations}. It also gives rise to photochemical reactivity.  Thus, it is unsurprising that vibronic coupling\textemdash absent from the JC and TC models\textemdash is a source of novel organic exciton-polariton behavior. The simplest way exciton-phonon coupling affects polariton behavior is by introducing an \textit{efficient} channel for nonradiative polariton decay (Table I). This happens because observed organic microcavity Rabi splittings are of the order of a few hundred meVs (Table I). In this energy interval, it is common for the molecular environment to have significant phonon DOS, which therefore plays an important role in assisting polariton relaxation.
Agranovich et al. first recognized in a seminal work \cite{agranovich2003} the effects of inhomogeneities on the organic polariton spectrum, and similarly, the role of resonant phonon emission and absorption on polariton relaxation dynamics. In particular, in Ref. \cite{agranovich2003}, the authors employed a macroscopic electrodynamics model to show that when the main source of disorder is inhomogeneous broadening of the molecular system (typically the case for organic microcavities), only the LP and UP states within a specific region of wave vector space (near the photon-exciton resonance) achieve large coherence lengths (typically a few $\mu m$) \cite{agranovich2003, aberra_guebrou_coherent_2012}. The remaining states with significant molecular character may be considered for practical purposes to form an incoherent reservoir [containing both the dark states described in Sec. \ref{jctcs} (Fig. \ref{jctc}) and also polaritons localized due to inhomogeneities] which is weakly-coupled to the cavity. Given that the latter are much more numerous than molecular polaritons, they form energy \textit{traps} fundamentally important in relaxation dynamics, as we discuss in Secs. \ref{orgrel}, \ref{orgappII}, and \ref{orgappIII}. Here we note that while the treatment in Ref. \cite{agranovich2003} is phenomenological, its fundamental conclusions were later confirmed by various numerical simulations and experimental data \cite{litinskaya2004, michetti_polariton_2005, litinskaya_loss_2006, agranovich_nature_2007, michetti_polariton_2008}. We discuss the relaxation dynamics of organic microcavities according to this picture in more detail in Sec. \ref{orgrel}.

An alternative approach to the investigation of organic cavity-polaritons was introduced by {\'C}wik et al. \cite{cwik2014}, who investigated their properties with a generalization of the Holstein Hamiltonian \cite{holstein1959} appropriate for the study of strongly coupled systems. In this \textit{Holstein-Tavis-Cummings} (HTC) model, the TC emitters (Eq. \ref{htc}) are assigned one or more independent vibrational degrees of freedom; these are linearly coupled to each organic exciton in accordance with the displaced oscillator model of vibronic coupling \cite{holstein1959} 
\ben H_{\text{exc-ph}} = \sum_{i=1}^N \sum_{j=1}^{N_{\text{ph}}}\lambda_j \hbar \omega_j \sigma_{+}^{(i)} \sigma_{-}^{(i)}(b_{ij}+b_{ij}^\dagger), \label{excph} \een
where the exciton operators follow the notation of the previous section, $b_{ij} (\omega_j)$  is the annihilation operator (natural frequency) of a harmonic phonon mode coupled to the $i$th-exciton, and $\lambda_j$ is the dimensionless vibronic coupling constant \cite{may2004charge}. Thus, in the absence of disorder, the (single-cavity mode) HTC Hamiltonian is given by

\begin{equation} H_{\text{HTC}} =  H_{\text{TC}} + H_{\text{ph}} + H_{\text{exc-ph}}, \label{htceq} \end{equation}
where $H_{\text{ph}} = \sum_{i=1}^N \sum_{j=1}^{\text{N}_{\text{ph}}} \hbar \omega_j b_{ij}^\dagger b_{ij}$ generates the free dynamics of $N_{\text{ph}}$ phonon modes per exciton.
The single-photon (exciton) eigenstates of Eq. \ref{htceq} (with a single phonon mode per molecule) were systematically investigated by Herrera and Spano in Refs. \cite{herrera2016, herrera_absorption_2017, herrera_dark_2017, herrera2018} (see also \cite{zeb2018, cwik_excitonic_2016}). These authors reported qualitatively distinct stationary states for $H_{\text{HTC}}$ depending on the ratio of Rabi splitting and phonon frequency $\Omega_R/\omega_v$. An important limit (with consequences discussed in Sec. \ref{orgappI} and \ref{orgappIII}) occurs when the light-matter interaction is much stronger than the local vibronic coupling, \textit{i.e.}, $\Omega_R/\lambda_v \omega_{v} \gg 1$. In this case, the phenomenon of \textit{polaron decoupling} is manifested \cite{spano_optical_2015, herrera2016}. This refers to a significant suppression of the vibronic coupling in the polariton states of a molecular ensemble strongly coupled to a microcavity (as discussed in Sec. \ref{jctcs}). It occurs as a consequence of the delocalized character of the polariton states (inherited from the photonic coherence volume and forced by the strong light-matter interaction); when the Rabi splitting is a few times larger than the considered vibronic couplings, the polaritons become (to a large extent) immune to the local (vibronic) perturbations acting on the excitonic states. This intuitive effect was studied long ago, as it is also the reason that delocalized excitations of organic J-aggregates have narrower lineshapes and weaker Stokes shift than the corresponding monomers \cite{knapp1984}. Further discussion of the different regimes of the HTC model is given in Refs. \cite{herrera2018, zeb2018}. It was also applied to the study of polariton effects on electron transfer \cite{herrera2016} (Sec. \ref{orgappI}), Raman spectrum \cite{strashko2016}, and organic polariton photoluminescence \cite{herrera_absorption_2017}. Notably, when vibrational relaxation and cavity leakage happen at comparable rates to the Rabi frequency \cite{herrera2018}, the behavior of the HTC eigenstates is essentially similar to that given by the theory first introduced by Agranovich et al. \cite{agranovich2003}. In this case, a simpler kinetic approach \cite{agranovich2003, herrera2018} where vibronic coupling acts a weak perturbation inducing incoherent scattering (see next Secs.) is well-suited to the description of organic polariton relaxation dynamics and photoluminescence. In particular, simulations of both phenomena are consistent with the LP being the main source of photoluminescence in microcavity experiments (though it was recently shown that with surface plasmons as the electromagnetic component, van Hove singularities arise and enable ultrafast photoluminescence from the UP \cite{yuen-zhou2017}). Given the timescales presented in Table I, the incoherent treatment of polariton-phonon dynamics is well-justified in many cases. Our further considerations will be based on it unless otherwise mentioned.

\subsection{Relaxation dynamics}\label{orgrel}
\begin{figure}
\centering
\includegraphics[scale=0.5]{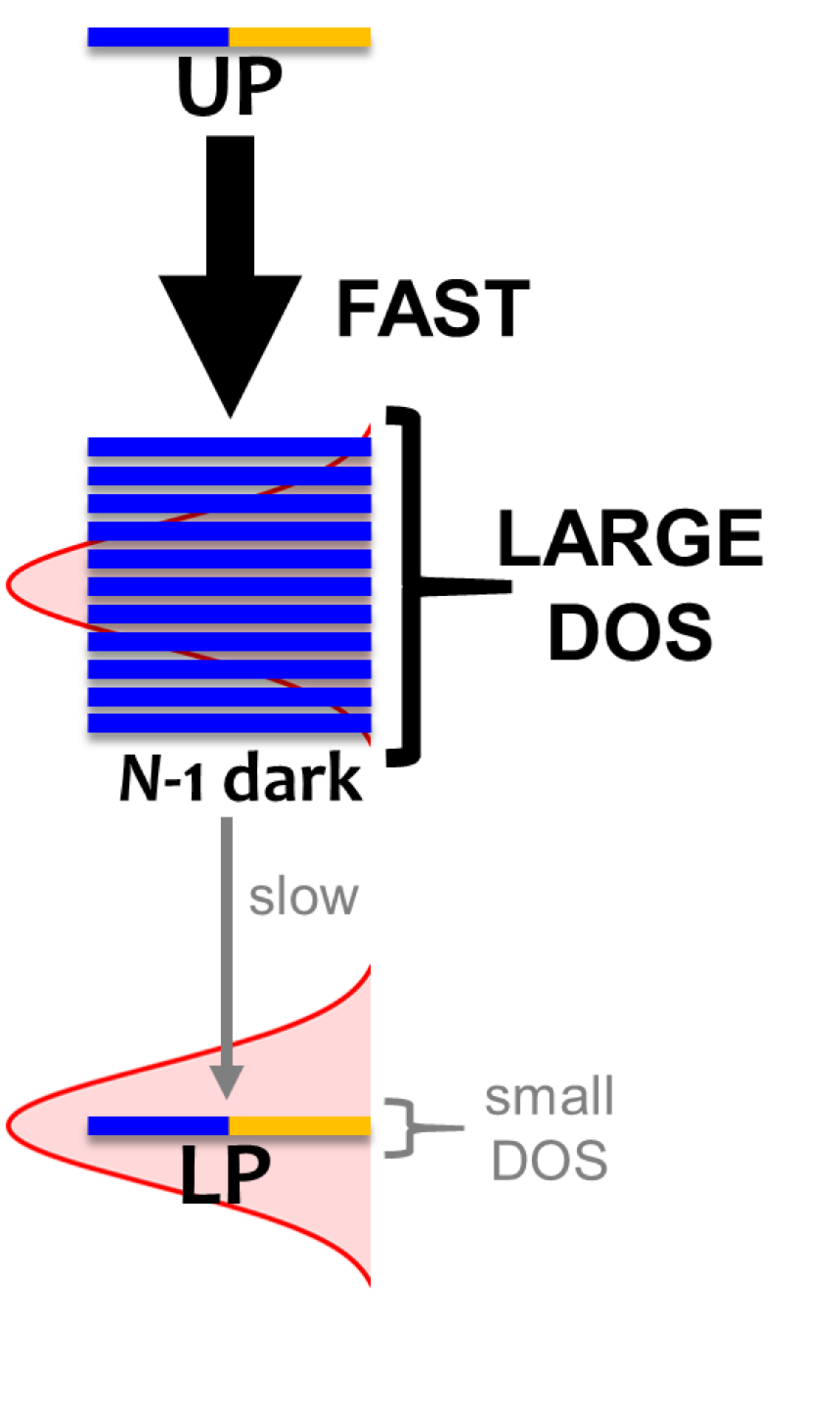}
\caption{Effect of DOS on vibrational-relaxation dynamics in the regime of
strong coupling between $N$ excitons and a single photon mode. For
large $N$, decay from UP to dark states (DOS $ \approx (N-1)/\text{exciton linewidth}$)
is much faster compared to that from dark states to LP (DOS $\approx 1/\text{LP linewidth}$) because transition rate scales with final DOS. When many modes are considered, the polariton bands have larger DOS but still much smaller than their dark state counterpart.}
\label{dosf}
\end{figure}
Given the lossy character of the microcavities and plasmonic layers routinely employed in in strong coupling experiments \cite{kavokin2017microcavities}, any practical use of organic polariton devices must account for the dissipative processes which may affect their performance. Typically, the damping of both bare cavity-photons and organic excitons can be reasonably approximated with a Markovian Master equation treatment \cite{nitzan2006chemical, gardiner1991quantum}. Said approach assumes these degrees of freedom interact weakly with a macroscopic bath characterized by its system-dependent spectral density \cite{nitzan2006chemical}. A choice needs to be made of whether each molecule has an independent bath, or a single set of environment modes interacts with all excitons. Both situations were explored in the work of del Pino et al. \cite{pino2015} (the pure-dephasing rates given in this work were corrected in Ref. \cite{martinez-martinez2018a}). Our discussion will assume the independent-baths scenario, which is the more realistic description for the study of disordered organic aggregates. 

The main dissipation channel for molecular polaritons involves the coupling of their photonic part to the external EM field modes via transmission through the cavity mirrors \cite{kavokin2017microcavities, savona1995}. This happens because most experiments employ optical resonators with low quality factor $Q(=\omega_c/\kappa$, where $\kappa$ is the cavity leakage rate  and also the full-width at half-maximum of the cavity mode of interest), thus leading to cavity-photon escape rates which are faster than molecular fluorescence or nonradiative decay (Table I). 

As mentioned, organic exciton-polaritons may also decay nonradiatively by vibronic coupling with molecular phonon modes (Table 1). Such relaxation occurs between polariton and dark states and is well described with Fermi's golden rule (FGR) \cite{dirac1927}. According to this framework, a quantum transition with higher density of final states will exhibit a faster rate compared to that with lower DOS if both processes are mediated by the same perturbations. The prominence of this DOS dependence in organic polariton relaxation dynamics was first characterized in Ref. \cite{agranovich2003}, which showed that via local phonon emission (Eq. \ref{excph}), the UP decays to the dark manifold much faster than the latter decays to the LP (Fig. \ref{dosf}). Agranovich et al. \cite{agranovich2003} considered a single Raman-active phonon-mode \cite{tartakovskii2001} with frequency nearly matching the Rabi splitting in Eq. \ref{excph} and inhomogeneously broadened spectral distribution of incoherent excitons for the dark band. The resulting vibrational-relaxation time to these ``dark" states (with one phonon) from the UP (with zero phonons and formed from exciton-resonant cavity-mode) was determined to be $\sim 50~\text{fs}$ (Table 1). This timescale is in good agreement with the low UP photoluminescence observed experimentally, given that the typical resolution of these measurements
is on the order of $100~\rm{fs}$ \cite{zhong2017}. In contrast,
a timescale of $\sim 10$ ps (Table 1) for the transition from the dark states to the LP band was obtained in Ref. \cite{litinskaya2004}.  Qualitatively, the difference between the rates of these relaxation processes is a direct manifestation of the final DOS in each case (Fig. \ref{dosf}). Indeed, even when considering light-matter coupling to the entire cavity-mode continuum (Sec. \ref{optspec}), the vast majority ($70-99\%$)\cite{agranovich2002, agranovich2003, litinskaya2004} of states with significant exciton character are dark/incoherent. Therefore, the latter form a reservoir which acts as an energy \textit{sink}. While inelastic scattering of dark modes may also increase the UP population, this process is relatively suppressed (timescale $\sim 1~ \mu\text{s}$; Table 1) as dictated by detailed balance. To further corroborate the association of vibrational relaxation with
photoluminescence, Michetti and LaRocca simulated organic microcavity emission with a kinetic model based on rates obtained with FGR \cite{michetti2008, michetti2009}. Experimental results were accurately reproduced, specifically the ratio of photoluminescence intensity of both polariton bands, as well as their temperature dependence \cite{michetti2008, michetti2009}. 

\subsection{Polariton-mediated photochemical reactivity}\label{orgappI}

The first report of drastic effects of polaritons on photochemistry was given by Hutchison et al. \cite{hutchison_modifying_2012}. In particular, a reduced rate was observed for  spiropyran-merocyanine photoisomerization under conditions where the \textit{product} of the transformation is resonant with the optical cavity. Later, Galego et al. \cite{galego2016} showed a mechanism for the suppression of polariton-mediated photochemical reactions where the \textit{reactants} are the strongly coupled species. In this case, the reaction rate decreases because the effective LP potential energy surface (PES) has a contribution from the (largely) non-reactive electronic ground-state PES (the reaction was assumed to proceed through the LP) \cite{galego2016}. Yet another example of polariton-mediated chemical reactivity was presented by Herrera and Spano \cite{herrera2016}. In this work, the regime of polaron decoupling (Sec. \ref{orgbas}) was assumed to show that nonadiabatic intramolecular electron transfer (ET) rates can be enhanced or suppressed when the electron-donor is strongly coupled with an optical cavity. A lower (higher) ET rate was shown to arise when the bare excited-donor and acceptor equilibrium geometries are displaced along the same (opposite) direction(s) relative to the electronic ground-state. In this case, the strong light-matter interaction induces a reduction (increase) of the difference between the electronically excited donor and acceptor equilibrium geometries, which effectively accelerates (inhibits) the reaction. Given that the energetics of the electronically excited-states determines the ET driving force, the manipulation of the polariton energies (Sec. \ref{jctcs}) provides yet another knob for the control of ET processes.

\subsection{Polariton-assisted remote long-range energy transfer}\label{orgappII}
\begin{figure}
\includegraphics[width=\columnwidth]{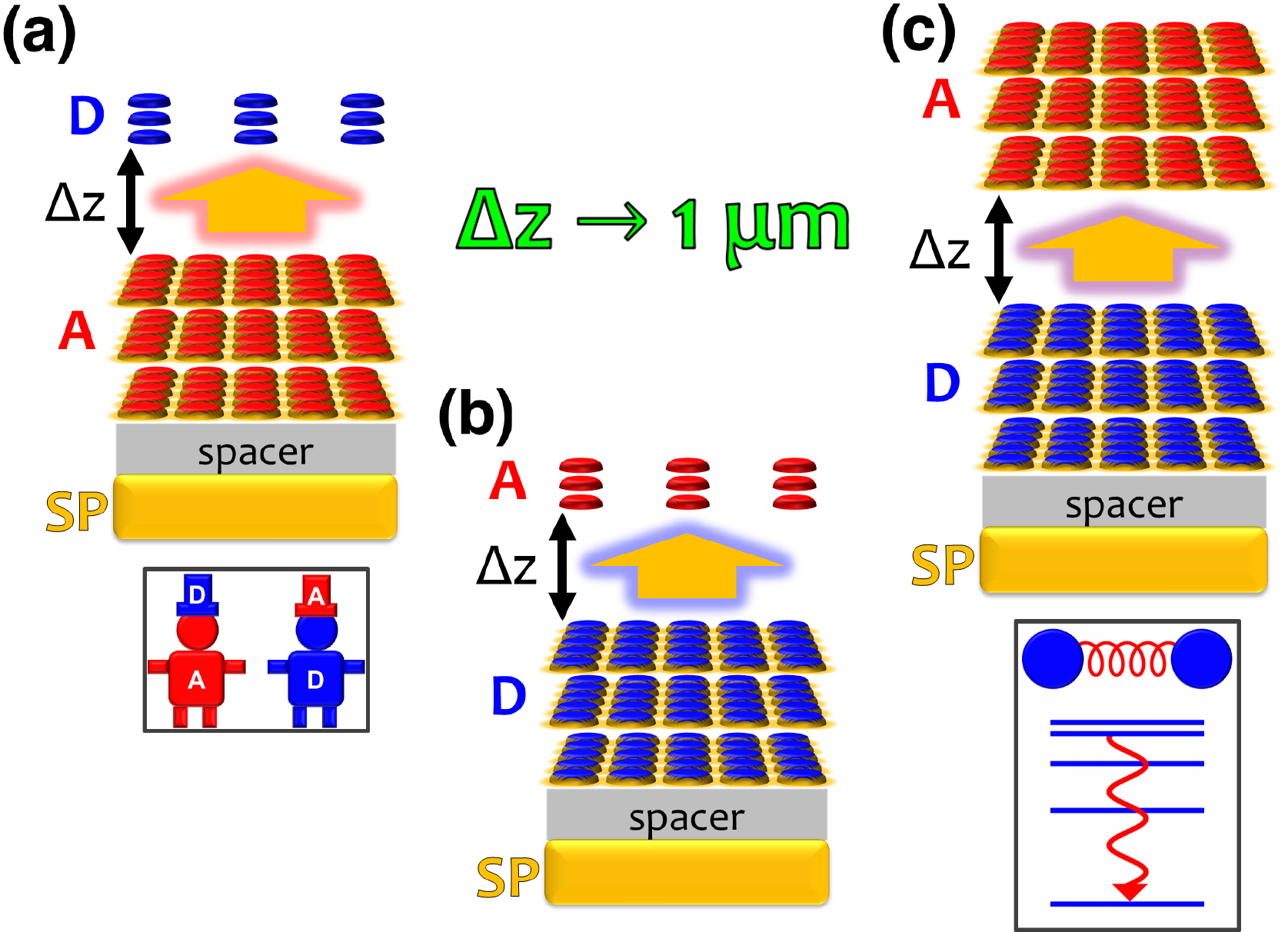}
\caption{Representations of polariton-assisted remote energy tranfer (PARET),
where donor-acceptor separation $\Delta z\approx 1\text{ }\mu\text{m}$ for various cases of strong coupling
to surface plasmons. (a) ``Carnival effect'' \textit{i.e.}, role-reversed)
PARET from dense slab of acceptors (featuring SC to SPs) to dilute
monolayer of donors. Inset: cartoon highlighting the ``carnival effect'',
or role reversal, between donors and acceptors. (b) PARET from dense
slab of strongly coupled donors to dilute monolayer of acceptors. (c) PARET from dense slab of donors to dense slab of acceptors (both are strongly coupled to SPs). Inset: cartoon illustrating the
vibrational relaxation that mediates PARET in this case. }
\label{fgr:paret}
\end{figure}
Excitation energy transfer (EET) converts the excitation of a donor (D) molecular species into that of a resonant acceptor (A) species \cite{may2004charge}. In most cases, this process is mediated by nonradiative dipole-dipole interactions and referred to as F\"orster resonance energy transfer (FRET) \cite{forster_zwischenmolekulare_1948}.  However, it is limited to molecular separations of $\sim 1-10~\text{nm}$ \cite{medintz2013fret}. Recently, it has been shown experimentally that efficient \textit{long-range} EET can be achieved in organic microcavities at the strong coupling regime \cite{coles2014b, zhong2017, georgiou2018}. A variant of this process was first studied by Basko et al. \cite{basko2000}, who investigated the effects of acceptor strong coupling on the decay of weakly-coupled donor excited-states without emphasis on the distance-independent character of the microcavity energy transfer. In a recent work \cite{du2017}, we provided a comprehensive theory of this phenomenon, which was denoted by \textit{polariton-assisted remote energy transfer} (PARET, Fig. \ref{fgr:paret}). The setup included separated donor and acceptor molecular slabs placed above a plasmonic layer. The distance-dependence of energy transfer rates was examined for \textit{exclusive} donor or acceptor strong coupling, and also for the case where \textit{both} chemical species are strongly coupled to the plasmonic layer. The effective Hamiltonian we employed is a simple generalization of the previously discussed models; it is given by
\ben H=H_{\text{D}}+H_{\text{A}}+H_{\text{P}}+H_{\text{DA}}+H_{\text{DP}}+H_{\text{AP}}, \label{hparet}  \een
where the first two terms on the right-hand side are HTC Hamiltonians (Eq. \ref{htceq})\textemdash except modified to include spatial variations in the light-matter coupling\textemdash for donor and acceptors.
Similarly, the SPs are described by $H_{\text{P}}$, which contains terms that describe both coherent and lossy plasmon dynamics. The interaction part of Eq. \ref{hparet} includes the weak dipole-dipole coupling between donor and acceptor states ($H_\text{{DA}}$) and plasmon resonance energy transfer (PRET) between excitons and SPs ($H_{\text{DP}}, H_{\text{AP}}$) \cite{liu2007}. Each strong coupling scenario (whether only one or both molecular slabs are strongly coupled) is associated to a distinct partitioning of $H$ into a zeroth-order Hamiltonian $H_{0}$ and a (weak) perturbation $V$. For instance, when the donor is the only strongly coupled species (Fig. \ref{fgr:paret}b), $H_{0} = H_{\text{A}} + H_{\text{D}} + H_{\text{P}}+H_{\text{DP}}$ and $V = H_{\text{DA}}+H_{\text{AP}}$. Given the partitioning appropriate for each scenario, the EET rates are obtained with FGR. In this case, EET from donor polaritons to bare acceptors was theoretically
predicted to happen even at micron donor-acceptor separations \cite{du2017}. Such PARET is attributable to the PRET contribution, which evanescently decays from the metal surface across distances as long as microns depending on the wave vector of the resonant SP. In contrast, the EET rate from the purely excitonic (donor) dark states to acceptors approaches that obtained in bare FRET for a sufficiently thick donor slab, as intuitively expected from a dense set of purely excitonic states. Conversely, strong coupling to only acceptors actually leads to a donor-to-polariton rate that is significantly smaller than the bare FRET. In analogy to our discussion of relaxation dynamics in Sec. \ref{orgrel}, this arises because the polariton band onto which the transfer is expected to happen has a much lower DOS than the dark state manifold. Furthermore, as in donor-exclusive strong coupling, the donor-to-dark-acceptors EET rate converges to the bare FRET rate (for a sufficiently thick acceptor slab). However, for intense enough acceptor-SP coupling, the donors and acceptors actually reverse roles (``carnival effect", Fig. \ref{fgr:paret}a) \cite{du2017}.
In contrast, in a different regime where strong coupling is realized with both donors and acceptors (Fig. \ref{fgr:paret}c), long-rate EET is mediated by vibrational relaxation \cite{du2017}. This induces transitions among polaritons\textemdash delocalized across donors and acceptors\textemdash and dark states with common excitonic character. By the same DOS arguments just discussed, EET to polaritons is much slower than that to the dark state manifolds. Nevertheless, the former is calculated to outcompete fluorescence, and the latter occurs as fast as molecular vibrational relaxation \cite{du2017}. Consequently, PARET from a mainly-donor to a mostly-acceptor state is theoretically attainable for chromophoric-slab separations of at least hundreds of nm. In fact, the computed rates for this case are in qualitative agreement with experimental data, even when the nature of the electromagnetic modes differ from one study to the other \cite{coles2014b}. 

It is worth mentioning that other schemes have been theoretically proposed to enhance excitation energy transport by exploiting strong light-matter coupling \cite{feist_extraordinary_2015, schachenmayer_cavity-enhanced_2015} (in conjunction with novel methods of topological protection \cite{yuen-zhou2014, yuen-zhou_plexciton_2016}).

\subsection{Polariton-assisted singlet fission}\label{orgappIII}
\begin{figure}
\includegraphics[width=\columnwidth]{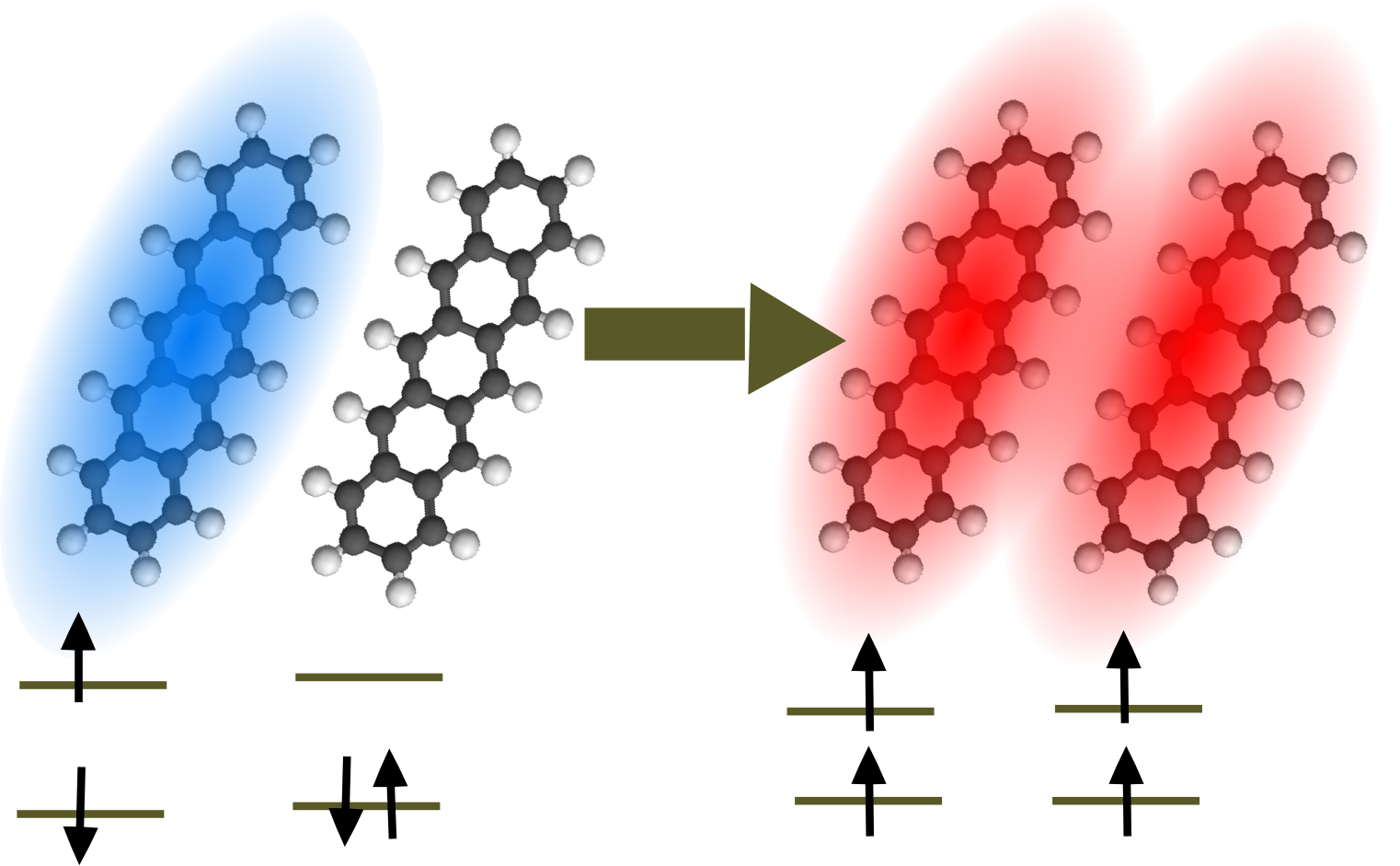}
\caption{Pictorial representation of singlet fission in pentacene. Blue (red) denotes a singlet (triplet) exciton.}\label{sf}\end{figure}

Singlet fission (SF) is a spin-allowed process where a (one-body) singlet exciton is converted into a (two-body) triplet-triplet (TT) state with vanishing total spin (Fig. \ref{sf}) \cite{smith2010, smith2013}. This phenomenon is of fundamental importance to the energy sciences, as it has been proven to enhance the efficiency of organic solar cells \cite{congreve2013, yost2014} by increasing the number of excitons produced per photon absorbed by an organic photovoltaic device, \textit{i.e.}, the external quantum yield (EQY). Given the demonstrated ability of molecular polaritons to influence chemical dynamics, it is natural to enquire what possibilities exist for the control of singlet fission in organic microcavities.

\begin{figure}
\includegraphics[width=\columnwidth]{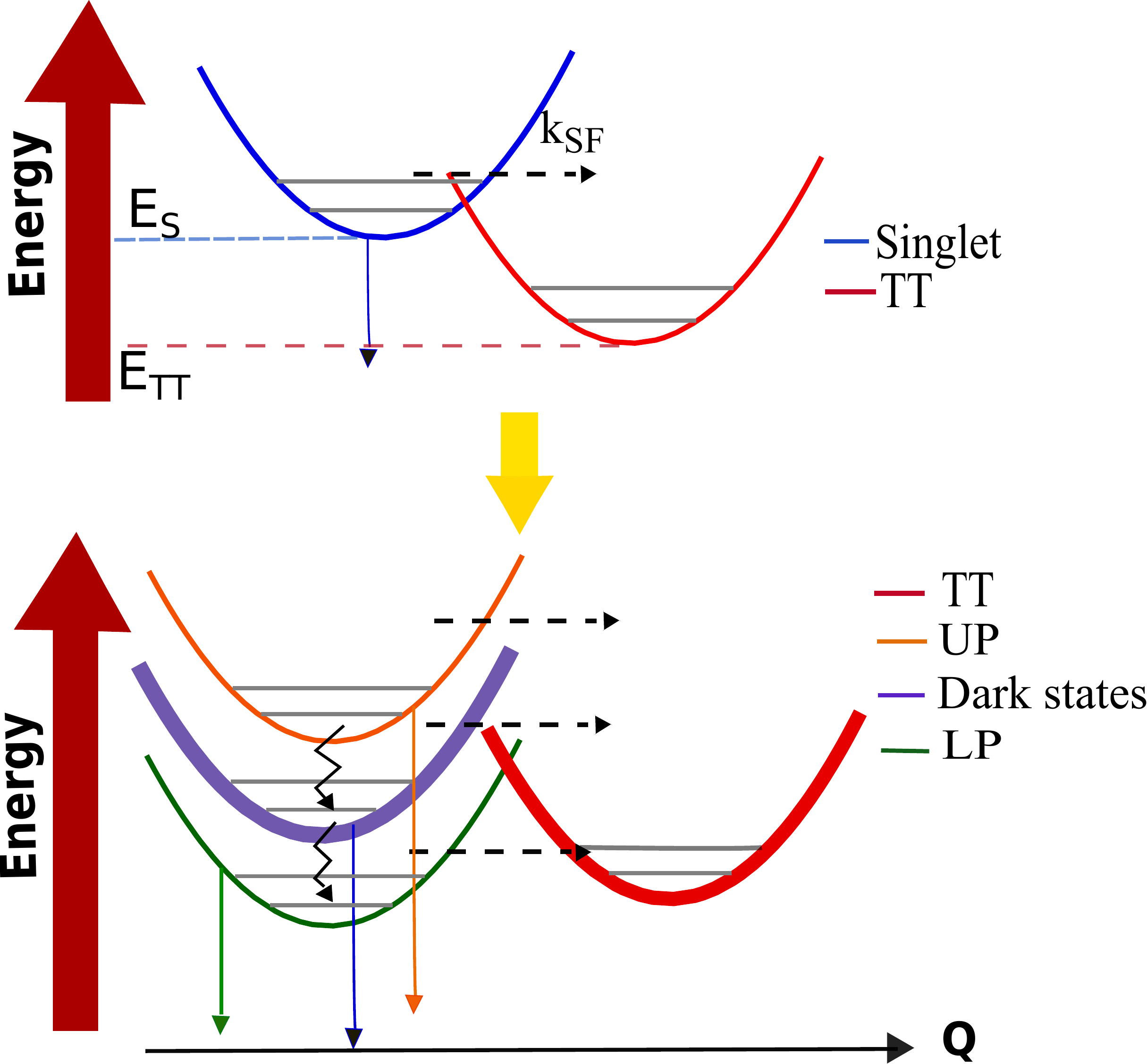}
\caption{Scheme of the transfer processes relevant to SF under normal
(top) and strong coupling (bottom) conditions. Solid (jagged) arrows indicate radiative (nonradiative) decay processes. Dashed arrows account for transitions between states with different electronic character. Thicker lines indicate larger DOS.} \label{sff}
\end{figure}

In Ref. \cite{martinez-martinez2017b}, we proposed a model for the investigation of polariton-assisted SF of acene chains in a microcavity which, for comparison purposes, also considered the competition of SF with other singlet quenching mechanisms. In order to quantitatively establish the effects of strong coupling on TT yield, Mart\'inez-Mart\'inez et al. employed the Pauli master equation formalism \cite{nitzan2006chemical, may2004charge}. The results highlight again (Secs. \ref{orgrel} and \ref{orgappII}) the essential (Fig. \ref{dosf}) influence of strong coupling on the DOS of donor and acceptor manifolds: the polariton manifold has a small DOS in comparison to the dark and TT. As a consequence, polariton decay to either dark or TT states is significantly faster than the reverse process. Another important finding is that to achieve polariton-based enhancement in the TT yield of an arbitrary SF material, the ideal candidate must have $\Delta G_{\text{SF}} = E_{\text{TT}}-E_{\text{S}}\ll0$ (see Fig. \ref{sff}). In this way, for sufficiently large Rabi splitting, the LP can be tuned close to resonance with
a high-frequency bath mode (known as the inner-sphere
in Marcus theory literature \cite{jortner1976}) of the TT states. This
reduces the energy barrier between the donor (LP) and acceptor states with respect to the bare material. Moreover, detailed balance implies thermal suppression of vibrational relaxation upward from LP to dark states at large Rabi splittings, and the most favorable decay channel directs singlet excited-state population to the TT manifold. In summary, Ref. \cite{martinez-martinez2017b} indicates that under experimentally accessible conditions polariton-assisted SF can outcompete SF quenching mechanisms, and turn materials with poor EQY into highly-efficient sensitizers.

\section{Vibrational-polaritons}\label{vibp}
\begin{figure}[h]
 \centering
\includegraphics[scale=0.35]{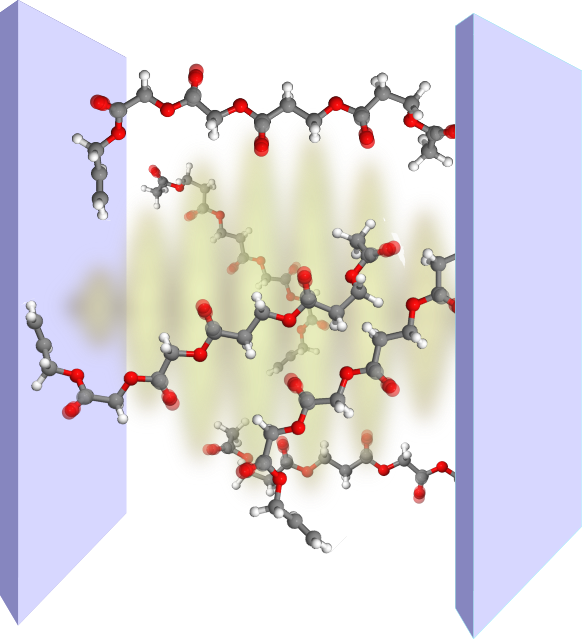}
\caption{Representation of strong coupling between a planar optical cavity and the carbonyl bonds of polyvinyl acetate chains.}\label{ircavf}\end{figure}

Vibrational-polaritons occur when dipole-active molecular vibrations interact strongly with the EM field of a microcavity (Fig. \ref{ircavf}). Studies of these novel excitations are stimulated by the possibilities they may offer for the \textit{selective} control of chemical bonds. In particular, there exists interest in employing vibrational strong coupling (VSC) to e.g., catalyze or inhibit chemical reactions \cite{thomas_ground-state_2016}, suppress or enhance intramolecular vibrational relaxation, and control the nonlinear optical response of molecular systems in the infrared (IR) \cite{xiang2017, dunkelberger2018b}. Furthermore, vibrational-polaritons might also provide desired novel sources of coherent mid-IR light.

Solid-state phonon-polaritons have been investigated since the 1960s \cite{henry1965, mills1974}, and some early studies of liquid-phase molecular vibrational-polaritons date back to the 1980s \cite{denisov1987}. However, it is only recently that \textit{cavity (or surface-plasmon) vibrational-polaritons} have been observed and systematically studied in the polymeric, liquid, and solution-phases \cite{shalabney_coherent_2015, george_liquid-phase_2015, long_coherent_2015, muallem2016a, muallem2016, simpkins_spanning_2015, casey_vibrational_2016, vergauwe_quantum_2016, thomas_ground-state_2016, dunkelberger_modified_2016, ahn2018, kapon_vibrational_2017, memmi2017, crum2018}. This is important because many important chemical reactions happen in the liquid-phase. Notably, under illumination with weak fields, the response of vibrational microcavities is similar to that of organic exciton-polaritons \cite{fano1956}. However, fundamentally novel behavior of vibrational-polaritons can be observed when higher excitations \cite{mukamel1999principles, hamm2011concepts} of the hybrid system are optically \cite{dunkelberger_modified_2016, xiang2017, ribeiro2017c, dunkelberger2018b}  or thermally \cite{thomas_ground-state_2016, chervy2018} accessed. In this Sec. we provide an overview of the properties of vibrational-polaritons with emphasis on their features which are qualitatively distinct from those of exciton-polaritons. We review the basic theory of VSC in Sec. \ref{vib_bas}, and discuss recently reported experimental and theoretical results on the \textit{nonlinear} interactions of vibrational-polaritons in Sec. \ref{vib_trans}. We conclude our discussion of IR strong coupling with some comments on recent tantalizing experimental observations of non-trivial VSC effects on chemical reactivity and IR emission which have been reported by the Ebbesen group \cite{thomas_ground-state_2016, chervy2018}.

\subsection{Basic features of vibrational strong coupling}\label{vib_bas}

In contrast to the electronic, the dynamics of a bare vibrational degree of freedom can be well-approximated at low energies by a \textit{weakly} anharmonic oscillator \cite{herzberg1939molecular}. This implies, e.g., that the $v = 1\rightarrow v = 2$ vibrational transition frequency $\omega_{12}$ is only weakly detuned from $\omega_{01}$, and the \textit{effective} transition dipole moment $\mu_{12}$ can be expressed as $\mu_{12} = \sqrt{2}\mu_{01}(1+\beta)$ where $\beta$ is typically a small number. However, these anharmonic properties can only be manifested in experiments that probe the \textit{nonlinear} \cite{mukamel1999principles, yuen2014ultrafast} optical response of vibrational-polaritons (Sec. \ref{vib_trans}). Still, there are important differences between the linear optical response of vibrational- and organic exciton-polaritons. For instance, while the excitonic strong coupling of organic aggregates is facilitated by their large transition dipole moments (e.g., $\mu_{01} \approx 5-15~\rm{D}$ in the case of J-aggregates \cite{valleau2012}), the intensity of vibrational transitions is often much weaker in comparison (in general $\mu_{01} <1.5~\rm{D}$). Thus, the Rabi splittings of vibrational-polaritons ($5-20$ meV \cite{shalabney_coherent_2015, long_coherent_2015, george_liquid-phase_2015, casey_vibrational_2016, thomas_ground-state_2016, vergauwe_quantum_2016, crum2018, memmi2017}) are generally weaker than those of organic microcavities (Table 1). However, vibrational linewidths are often much smaller compared to those of organic excitons. In addition, resonant IR microcavities have lower photon leakage rates ($0.1-5 ~\text{ps}$\cite{shalabney_coherent_2015, long_coherent_2015, george_liquid-phase_2015, casey_vibrational_2016, thomas_ground-state_2016, vergauwe_quantum_2016, crum2018, memmi2017}) than the organic (Table 1), for the wavelength of vibrational transitions generally belongs to the mid-IR ($\lambda = 3-30~\mu m$) \cite{kavokin2017microcavities}. Thus, there exist many opportunities for strong coupling of cavity EM fields with molecular vibrational degrees of freedom. Typically, molecular vibrations with large absorptivity are dominated by polar functional groups such as carbonyl (C$=$O), amide ($\text{H}_2\text{N}-\text{C$=$O}$) and cyanide ($\rm{C}\equiv N$). In fact, most of the observed vibrational-polaritons arose from the strong coupling of IR cavities with the  $\rm C=O$ or $\rm C\equiv N$ bonds of organic polymers \cite{shalabney_coherent_2015, long_coherent_2015, muallem2016}, neat organic liquids \cite{george_liquid-phase_2015}, polypeptides \cite{vergauwe_quantum_2016}, transition metal complexes \cite{casey_vibrational_2016, simpkins_spanning_2015}, and liquid crystals \cite{hertzog2017}. Yet, given the dependence of the collective Rabi splitting on the molecular density, there is no requirement that the strongly coupled bonds need to be significantly polarizable; in fact, vibrational-polaritons have been also been reported for alkene ($\text{C$=$C}$) \cite{george_liquid-phase_2015}, and silane $\text{C$-$\text{Si}}$ \cite{thomas_ground-state_2016} bonds.
Based on the discussion above, the \textit{harmonic} Hamiltonian describing VSC of a lossless single-mode IR cavity with an ensemble of $N$ independent \textit{identical} molecular vibrational modes is given by 
\begin{equation} H^{(0)} = \hbar \omega_c a^\dagger a + \hbar\omega_0 \sum_{i=1}^N b_i^\dagger b_i - \hbar g_s \sum_{i=1}^N \left(b_i^\dagger a  + a^\dagger b_i\right), \label{h0v} \end{equation}
where $b_i$ ($b_i^\dagger$) denotes the \textit{bosonic} annihilation (creation) operator for the vibration localized at molecule $i$, and the other constants are defined in Sec. \ref{bas}. A more realistic description of the system would include the dissipative dynamics  of both cavity and matter degrees of freedom. However, here simplifications arise relative to the description of organic exciton-polaritons: vibrational spectra show no Stokes shift, and their absorption bands are in some cases dominated by homogeneous broadening. Thus, it is in general easier to model the effects of cavity and vibrational damping on the polariton linewidths. In particular, the vibrational environment (represented by both intra and intermolecular degrees of freedom) may be accurately modeled as a thermal distribution of harmonic oscillators (bath) which interact weakly with the system \cite{nitzan2006chemical}. It is again reasonable to assume that the bath of each vibrational degree of freedom is independent (Sec. \ref{orgrel}). Under these conditions and the usual assumptions of dissipative Markovian dynamics \cite{gardiner1991quantum, may2004charge}, it can be shown that the IR cavity optical response is determined by the normal mode frequencies and dissipation rates of the classical problem of two coupled damped oscillators representing the cavity photon and the bright (totally-symmetric) superposition of molecular excited-states (see e.g., the Supporting Information of \cite{ribeiro2017c}). 
 
\subsection{Transient vibrational-polaritons}\label{vib_trans}
 
\begin{figure}[h]
  \centering
    \includegraphics[scale=0.5]{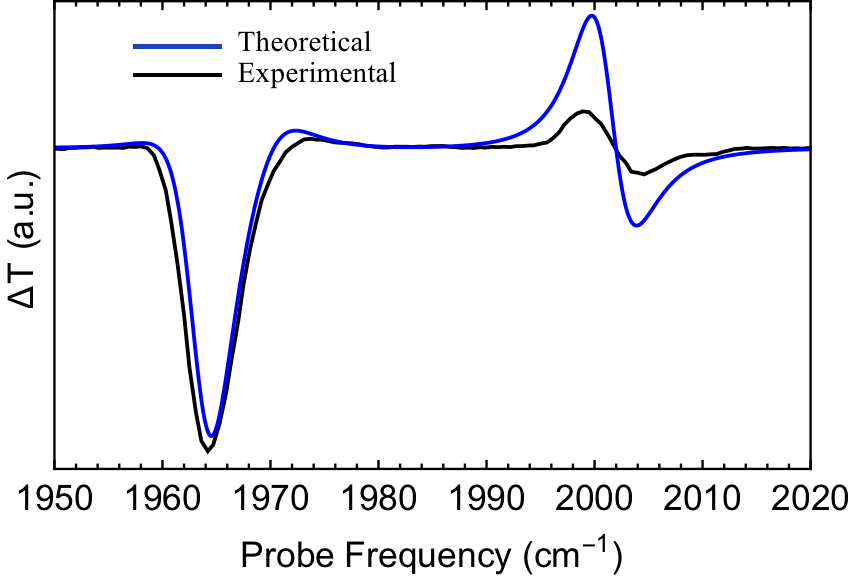}
    \caption{Experimental and theoretical pump-probe (differential) transmission spectra of strongly coupled $\text{W(CO)}_6$ in an optical microcavity \cite{ribeiro2017c}. These results correspond to the case where the cavity photon and the molecular vibration (asymmetric C$=$O stretch) are resonant.} \label{trvib}     
  \end{figure}

\begin{figure}
  \centering
\includegraphics[scale=0.4]{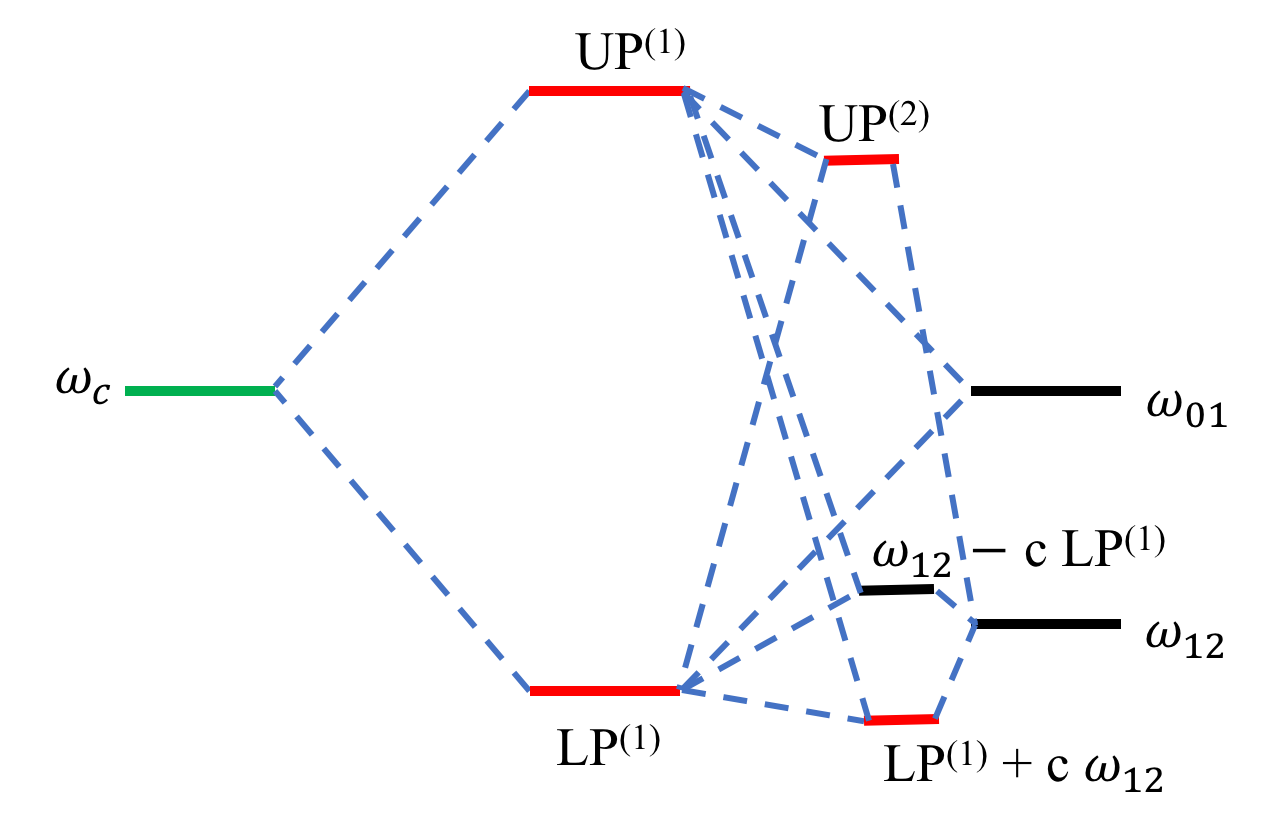}
\caption{Energy level hybridization diagram including the photonic ($\omega_c$) and vibrational transitions ($\omega_{01}, \omega_{12}$) involved in the formation of linear $[\text{LP}^{(1)},\text{UP}^{(1)}]$ and (effective) transient vibrational-polaritons $[\text{UP}^{(2)}$, and the combination of $\text{LP}^{(1)}$ and the  $1\rightarrow 2$ vibrational transition] \cite{xiang2017}. Note the significant interaction between the linear LP mode and the $1\rightarrow 2$ excitations (represented by the coefficient $c$) arising from the incoherent population of vibrational modes induced by a pump after sufficiently long probe-delay times (see text).} \label{orbf} 
\end{figure}

The first pump-probe (pp) spectra of vibrational-polaritons were obtained by Dunkelberger et al. \cite{dunkelberger_modified_2016}. These experiments employed liquid-phase solutions of W(CO)$_6$ in hexane. The T$_{1u}$ triply-degenerate carbonyl mode was chosen to couple to the cavity as its effective transition dipole moment is relatively large ($\approx 1\rm{D}$), and its linewidth is sufficiently small ($\approx 3 ~\text{cm}^{-1}$). Further insight on the nonlinear behavior of $\text{W(CO)}_6$ vibrational-polaritons was reported recently by Xiang et al. \cite{xiang2017}, who also provided the first 2D spectra of vibrational-polaritons. Both the pp and 2D spectra showed unambiguous evidence of \textit{asymmetric} polariton\textemdash polariton and polariton\textemdash dark state interactions \cite{dunkelberger_modified_2016, xiang2017}, \textit{even} when linear reflectivity measurements showed equally-intense LP and UP response. In particular, the IR cavity differential probe transmission (pp transmission minus linear probe transmission) displayed a consistently large (small) negative feature at the linear LP (UP) frequency, and positive shifted (relative to the linear spectrum) transmission resonances for the LP and UP (Fig. \ref{trvib}). These observations were interpreted in Ref. \cite{ribeiro2017c} with a microscopic model of vibrational-polaritons which included the effect of vibrational anharmonicity on the polariton optical response. Both \textit{mechanical} and \textit{electrical} nonlinearities were added to the model described by Eq. \ref{h0v}.  Mechanical (or bond) anharmonicity represents the tendency that bonds break at high energies, while electrical anharmonicity occurs due to nonlinearity of the effective vibrational transition dipole moment with respect to  small displacements of the nuclei from equilibrium (e.g., due to non-Condon effects \cite{gerhard1968infrared, khalil_coherent_2003, ishii2009}). In practice, the main effect of mechanical and electrical nonlinearities is to redshift overtone transitions from the fundamental and give band absorption intensities which violate the harmonic oscillator scaling, respectively. Thus, the effective Hamiltonian of anharmonic vibrational-polaritons interacting with a single cavity-mode can be written as \cite{ribeiro2017c}
\begin{equation} H = H^{(0)} - \hbar \alpha \sum_{i=1}^N b_i^\dagger b_i^\dagger b_i b_i -\hbar \beta \sum_{i=1}^{N}\left(b_i^\dagger b_i^\dagger b_i a +a^\dagger b_i^\dagger b_i b_i \right), \label{hv} \end{equation}
where $\alpha$ characterizes mechanical anharmonicity, \textit{i.e.}., $2\alpha = \omega_{12}-\omega_{10}$, and $\beta$ parametrizes the deviation of $\mu_{12}$ from that predicted for a harmonic dipole. This theory provided pp spectra with the same essential features as experimentally reported (Fig. \ref{trvib}) \cite{xiang2017, ribeiro2017c}. It shows that the pump-probe transmission contains three resonances resulting from the interaction of cavity photons with a population of molecular vibrations in the ground and first excited-states (the latter of which is a byproduct of the pump excitation of the system at earlier times). The largely suppressed probe-transmission (large negative signal in Fig. \ref{trvib}) in a neighborhood of the linear LP frequency is a result of its near-resonance with the $1 \rightarrow 2$ transition of dark states. The effect of the nonlinearity is weaker in UP since its frequency is highly off-resonant with $\omega_{12}$ (Fig. \ref{orbf}). Given that vibrational anharmonicity generally manifests as $\omega_{12} < \omega_{01}$, the much larger
anharmonicity of LP (compared to UP) is expected to be a \textit{generic} feature of IR microcavities. In other words, the studies discussed in this Sec. indicate that vibrational-LP modes are \textit{softer} than the UP. Further corroboration of this theory came in a recent study by Dunkelberger et al. \cite{dunkelberger2018b} who measured the pump-probe spectra of the $\text{W(CO)}_6-\text{hexane}$ system at low concentrations such that the Rabi splitting was small enough for the LP to be off-resonant with $\omega_{12}$ by nearly  $10~\text{cm}^{-1}$ (in this case, given that the LP and UP are both significantly off-resonant with $\omega_{12}$, the asymmetry in the transient polariton response is diminished).

\subsection{Applications: Chemical kinetics and thermal emission}\label{vib_app}

We conclude our discussion of vibrational-polaritons by mentioning two recent observations of the effects of VSC on chemical reactions and thermal emission. First, Thomas et al. \cite{thomas_ground-state_2016} provided conclusive evidence that an organic silane deprotection reaction proceeds via a different mechanism under conditions where the C$-$Si bond is strongly coupled to an optical microcavity, \textit{even in the absence of external photon pumping of the polariton system}. Specifically, the reaction rate was measured as a function of temperature under normal and VSC conditions, and the resulting kinetic curves provide transition-state theory estimates for the entropy and enthalpy of activation. The entropy of activation was reported to be positive under VSC, but negative otherwise.  In addition, the kinetics was strongly dependent on the Rabi splitting and, e.g., under weak coupling, the reaction rate was indistinguishable from that measured outside the cavity. Similarly puzzling results were shown recently by Chervy et al. \cite{chervy2018}, who reported non-thermalized thermal emission of cavity ($\text{C}=\text{O}$)  vibrational-polaritons of an organic polymer at $373 ~\text{K}$. It was also observed that while the bare polymer and cavity emission spectra matched the theoretical thermal emission, the strongly coupled system showed emission peaks at frequencies displaced from the expected (based on the linear optical spectra). 

These experiments show the rich dynamics featured by vibrational-polaritons. They have in common the fact that both investigated phenomena arise from thermally-activated \textit{anharmonic} excited-state dynamics of vibrational-polaritons. Further work is needed to understand the sources of the observed behaviors. We expect that their microscopic interpretation will likely shed light on novel ways to control chemical bonds with VSC.

\section{Ultrastrong coupling}\label{ultra}

All of our previous considerations assumed that the (collective) Rabi splitting was stronger than the dissipative couplings of the bare molecule (or cavity), but also much weaker than the transition energy of interest. The ultrastrong coupling (USC) regime is characterized by the violation of the latter assumption \cite{ciuti_quantum_2005, ciuti_input-output_2006}.
In particular, the onset of USC is conventionally defined to arise for vacuum Rabi splittings that satisfy $\Omega_R/{\omega_0} > 10\%$ \cite{moroz2014, ciuti_quantum_2005}. When this condition is fulfilled, significant deviations from the approximate light-matter coupling assumed in Eqs. \ref{jceq} and \ref{tceq} become relevant. In particular, at USC, states with different excitation number (number of photons + molecular excited-states) are allowed to hybridize, while our previous discussion assumed that the interaction of radiation with bright-molecular states preserves the total excitation number of the systems. An essential consequence is that while the ground-state of a strongly coupled system $(\ket{0})$ is indistinguishable from the decoupled where all degrees of freedom are at their ground-state, the lowest-energy state of an ultrastrongly coupled system is a superposition of states consisting of correlated photons and delocalized bright molecular excitations \cite{ciuti_quantum_2005, ciuti_input-output_2006}. Notably, molecular USC was first achieved less than ten years ago \cite{schwartz2011}. While this field has seen considerable progress including recent reports of organic exciton \cite{schwartz2011, kena-cohen2013, balci2013, gambino2014, george_ultra-strong_2015} and vibrational USC \cite{george_multiple_2016}, the exploration of USC effects on chemical transformations is only beginning to be understood. We discuss a specific case below. 

\begin{figure}
 \centering
\includegraphics[width=0.75\columnwidth]{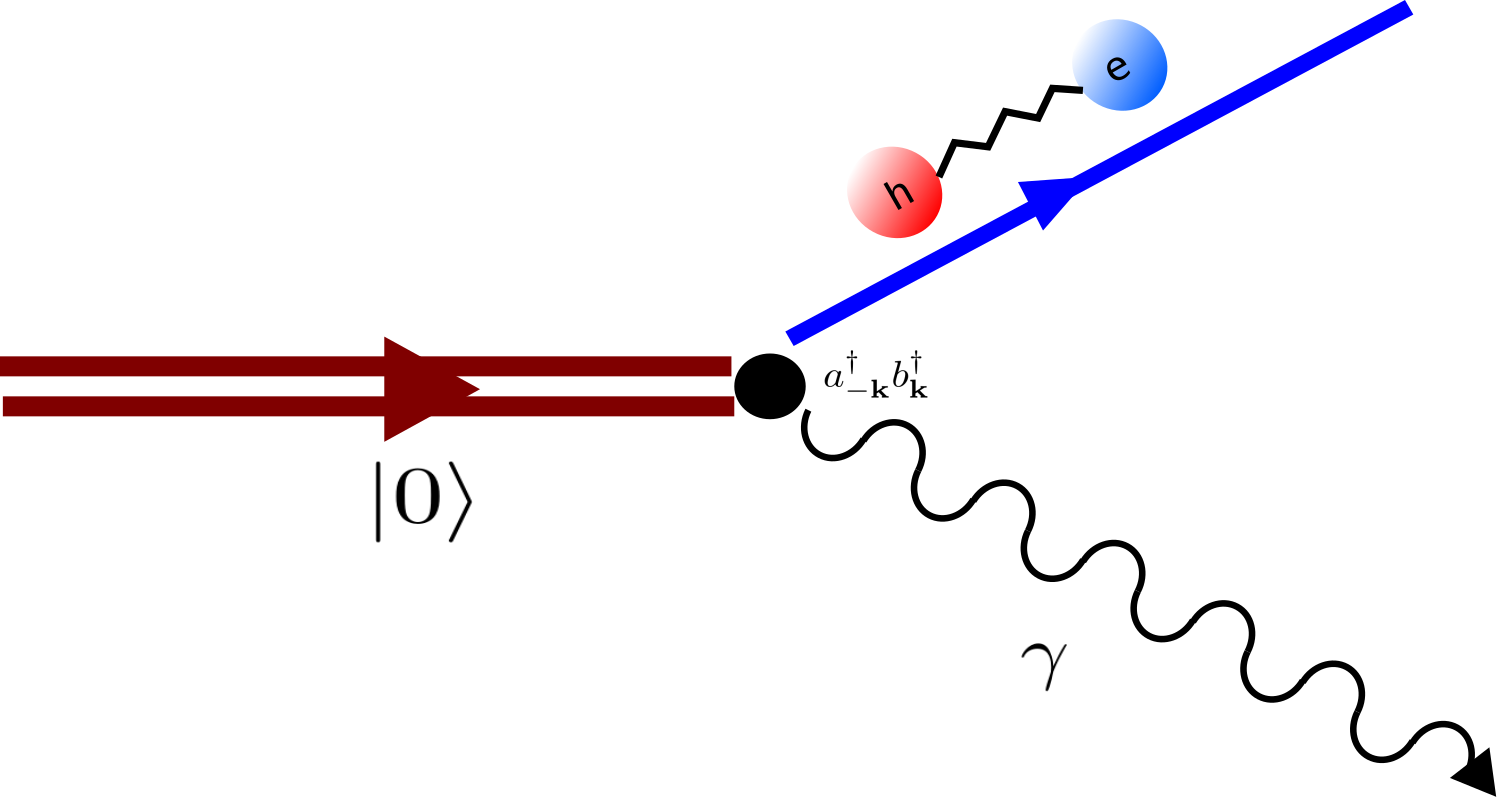}
\caption{Feynman diagram for spontaneous production of correlated exciton-photon pairs from the bare system ground-state $\ket{0}$. This  process is significant in the ultrastrong coupling regime where light-matter couplings of the form $(a_{-\mathbf{k}}^\dagger b_{\mathbf{k}}^\dagger + \hc)$ become relevant (see Eq. \ref{ple-u})}\label{ultraf}\end{figure}

In a recent work \cite{martinez-martinez2018}, we studied the effects of USC in the electronic ground-state energy
landscape of a molecular ensemble. In particular, we considered a simplified model of a molecular slab interacting with a plasmonic field in the USC regime. The Born-Oppenheimer Hamiltonian
of the system is given by
\ben H_{\text{BO}} = T_{\text{nuc}}+H_{\text{el}}(\mathbf{R}), \een where $T_{\text{nuc}}$ is
the total nuclear kinetic energy operator, $\mathbf{R}$ denotes the nuclear configuration of all molecules, and
\ben H_{\text{el}}(\mathbf{R})= H_{\text{g}}(\mathbf{R}) + H_{\text{pl}}+H_{\text{e}}(\mathbf{R})+H_{\text{pl-e}}(\mathbf{R}), \label{hult} \een
where $H_{\text{g}}(\mathbf{R})=\sum_{\mathbf{n}}\hbar \omega_{g}(R_{\mathbf{n}})$ is the Born-Oppenheimer electronic ground-state energy of the ensemble at
an arbitrary nuclear configuration $\{R_{\mathbf{n}}\}$ ($R_{\mathbf{n}}$ is the
nuclear coordinate of molecule located at site $\mathbf{n}$ within
the molecular slab), $H_{\text{pl}}=\sum_{\mathbf{k}}\hbar \omega_{\mathbf{k}}a_{\mathbf{k}}^{\dagger}a_{\mathbf{k}}$ is the Hamiltonian of bare plasmon modes with dispersion $E(\mathbf{k}) = \hbar \omega_{\mathbf{k}}$ (where $\mathbf{k}$ is the plasmon in-plane wave vector) and creation (annihilation) operators $a_{\mathbf{k}}^{\dagger}$
($a_{\mathbf{k}}$). The exciton Hamiltonian for ensemble nuclear configuration $\{R_{\mathbf{n}}\}$ is given by $H_{\text{e}}(\mathbf{R})=\sum_{\mathbf{n}}\left[\hbar \omega_{e}(R_{\mathbf{n}})-\hbar\omega_{g}(R_{\mathbf{n}})\right]b_{\mathbf{n}}^{\dagger}(R_{\mathbf{n}})b_{\mathbf{n}}(R_{\mathbf{n}})$, where $b_{\mathbf{n}}^{\dagger}(R_{\mathbf{n}})$ $[b_{\mathbf{n}}(R_{\mathbf{n}})]$ is the creation (annihilation) operator of the $\mathbf{n}$th-site.
The exciton-plasmon interaction is given by 
\ben H_{\text{pl-e}}(\mathbf{R})=\sum_{\mathbf{k}}\sum_{\mathbf{n}}\hbar g_{\mathbf{k}}^{\mathbf{n}}(R_{\mathbf{n}})\left(a_{\mathbf{k}}^{\dagger}+a_{\mathbf{k}}\right)\left[b_{\mathbf{n}}^{\dagger}(R_{\mathbf{n}})+b_{\mathbf{n}}(R_{\mathbf{n}})\right], \label{ple-u} \een
where $g_{\mathbf{k}}^\mathbf{n}(R_{\mathbf{n}})$ is the interaction between the plasmon with wave vector $\mathbf{k}$ and the $\mathbf{n}$th exciton. It depends on the position and geometry of the molecule since the plasmonic electric fields vary in space and the molecular transition dipole moment is assumed to depend on $R_\mathbf{n}$. Notice the explicit inclusion in $H_{\text{pl-e}}$ of
terms that do not preserve the total number of excitations (see also Fig. \ref{ultraf}). We also note that the maximal value of the collective couplings obtained with this model does not surpass $20\%$ of the exciton gap; this justifies the neglect of the  EM field diamagnetic terms in Eq. \ref{hult} \cite{deliberato2014}. The molecules that constitute the referred ensemble can undergo isomerization. This is described
by the electronic ground and excited adiabatic PESs, $\hbar \omega_{g}(R_{\mathbf{n}})$ and $\hbar \omega_{e}(R_{\mathbf{n}})$,
respectively. The former has a double-well structure and an avoided-crossing with the latter. Ref. \cite{martinez-martinez2018} analyzed various cross-sections of the dressed (collective) ground-state PES arising under USC. This included the cut where all molecular coordinates were frozen at the reactant configuration except for a single molecule. Thus, such reaction coordinate represents an \textit{effective} single-molecule PES. However, it was observed that the effective reaction barrier is almost unaffected by the collective light-matter coupling. Rather, the maximal energetic shifts induced by USC were identified as Lamb shifts which are small in comparison to the thermal energy. 
However, these results do not discourage further application of the ultrastrong coupling to chemical systems. The conditions studied in Ref.  \cite{martinez-martinez2018} were such that the light-matter interaction was near the edge of the USC regime, where the total exciton (photon) population in the ground-state is small, and a perturbative treatment of their effects is valid. In this case, the USC ground-state deviation from the bare system is nearly inconsequential. We believe that future theoretical and experimental studies of USC in the non-perturbative regime will present novel possibilities for electronic ground-state chemical dynamics.

\section{Epilogue}\label{epi}
We hope to have convinced the reader that: (i) the phenomena emergent from the (ultra)strong coupling regime presents novel opportunities for the control of chemical transformations induced by electronic and vibrational dynamics, and (ii) there remains much experimental and theoretical work to be done to unravel all of the intricacies and possibilities of polariton-mediated chemistry. Future experimental work will certainly entertain creative ways to steer chemical events using optical cavities in various regimes of external pumping and thermodynamic conditions, as well as new opportunities to harness many-body quantum effects towards the control of physicochemical properties of molecules. From the theoretical perspective, we expect novel applications and further development of effective condensed matter theories that describe the emergent phenomenology afforded by molecular polaritons. As we have shown here, these theories are particularly powerful in predicting nontrivial thermodynamic-limit behavior which can be directly employed to guide experiments. Lastly, there is a push towards the development of \textit{ab initio} quantum chemistry and quantum and semiclassical dynamics methodologies to simulate molecular polaritonic systems with atomistic detail \cite{ruggenthaler2014, bennett2016a, flick2017, luk_multiscale_2017, vendrell2017coherent}. Future studies of molecular polariton theory are expected to integrate quantum optics with the standard toolbox of chemical dynamics including e.g., surface-hopping methods \cite{subotnik2016}, quantum master equations and path-integral approaches \cite{makri1995, tanimura2006}. Still, the complex interplay between electronic, nuclear, and photonic degrees of freedom in complex dissipative environments presents a whole new set of challenges for computational methods, which will require novel solutions. 

\section*{Conflicts of interest}
There are no conflicts to declare.

\section*{Acknowledgments}
RFR, MD, JCGA, and JYZ acknowledge support from the NSF CAREER Award CHE-164732. LAMM and JCGA are grateful for the support of the UC-Mexus
CONACyT scholarship for doctoral studies. All authors were partially supported by generous UCSD startup funds. We acknowledge illuminating discussions we had throughout our collaborations with Wei Xiong, Stephane Kena-Cohen, Vinod Menon, Jeff Owrutsky, Adam Dunkelberger, Blake Simpkins, and Bo Xiang.


\balance


\bibliography{lib} 

\providecommand*{\mcitethebibliography}{\thebibliography}
\csname @ifundefined\endcsname{endmcitethebibliography}
{\let\endmcitethebibliography\endthebibliography}{}
\begin{mcitethebibliography}{169}
\providecommand*{\natexlab}[1]{#1}
\providecommand*{\mciteSetBstSublistMode}[1]{}
\providecommand*{\mciteSetBstMaxWidthForm}[2]{}
\providecommand*{\mciteBstWouldAddEndPuncttrue}
  {\def\EndOfBibitem{\unskip.}}
\providecommand*{\mciteBstWouldAddEndPunctfalse}
  {\let\EndOfBibitem\relax}
\providecommand*{\mciteSetBstMidEndSepPunct}[3]{}
\providecommand*{\mciteSetBstSublistLabelBeginEnd}[3]{}
\providecommand*{\EndOfBibitem}{}
\mciteSetBstSublistMode{f}
\mciteSetBstMaxWidthForm{subitem}
{(\emph{\alph{mcitesubitemcount}})}
\mciteSetBstSublistLabelBeginEnd{\mcitemaxwidthsubitemform\space}
{\relax}{\relax}

\bibitem[Turro(1991)]{turro1991modern}
N.~J. Turro, \emph{Modern Molecular Photochemistry}, {University science
  books}, 1991\relax
\mciteBstWouldAddEndPuncttrue
\mciteSetBstMidEndSepPunct{\mcitedefaultmidpunct}
{\mcitedefaultendpunct}{\mcitedefaultseppunct}\relax
\EndOfBibitem
\bibitem[Balzani \emph{et~al.}(2014)Balzani, Ceroni, and
  Juris]{balzani2014photochemistry}
V.~Balzani, P.~Ceroni and A.~Juris, \emph{Photochemistry and Photophysics:
  Concepts, Research, Applications}, {John Wiley \& Sons}, 2014\relax
\mciteBstWouldAddEndPuncttrue
\mciteSetBstMidEndSepPunct{\mcitedefaultmidpunct}
{\mcitedefaultendpunct}{\mcitedefaultseppunct}\relax
\EndOfBibitem
\bibitem[Hopfield(1958)]{hopfield_theory_1958}
J.~J. Hopfield, \emph{Physical Review}, 1958, \textbf{112}, 1555--1567\relax
\mciteBstWouldAddEndPuncttrue
\mciteSetBstMidEndSepPunct{\mcitedefaultmidpunct}
{\mcitedefaultendpunct}{\mcitedefaultseppunct}\relax
\EndOfBibitem
\bibitem[Agranovich(1959)]{agranovich1960dispersion}
V.~M. Agranovich, \emph{Sov. Phys. JETP}, 1959, \textbf{10}, 307--313\relax
\mciteBstWouldAddEndPuncttrue
\mciteSetBstMidEndSepPunct{\mcitedefaultmidpunct}
{\mcitedefaultendpunct}{\mcitedefaultseppunct}\relax
\EndOfBibitem
\bibitem[Ebbesen(2016)]{ebbesen_hybrid_2016}
T.~W. Ebbesen, \emph{Accounts of Chemical Research}, 2016, \textbf{49},
  2403--2412\relax
\mciteBstWouldAddEndPuncttrue
\mciteSetBstMidEndSepPunct{\mcitedefaultmidpunct}
{\mcitedefaultendpunct}{\mcitedefaultseppunct}\relax
\EndOfBibitem
\bibitem[Kavokin \emph{et~al.}(2017)Kavokin, Baumberg, Malpuech, and
  Laussy]{kavokin2017microcavities}
A.~V. Kavokin, J.~J. Baumberg, G.~Malpuech and F.~P. Laussy,
  \emph{Microcavities}, {Oxford University Press}, 2017, vol.~21\relax
\mciteBstWouldAddEndPuncttrue
\mciteSetBstMidEndSepPunct{\mcitedefaultmidpunct}
{\mcitedefaultendpunct}{\mcitedefaultseppunct}\relax
\EndOfBibitem
\bibitem[Baranov \emph{et~al.}(2018)Baranov, Wers{\"a}ll, Cuadra, Antosiewicz,
  and Shegai]{baranov2018}
D.~G. Baranov, M.~Wers{\"a}ll, J.~Cuadra, T.~J. Antosiewicz and T.~Shegai,
  \emph{ACS Photonics}, 2018, \textbf{5}, 24--42\relax
\mciteBstWouldAddEndPuncttrue
\mciteSetBstMidEndSepPunct{\mcitedefaultmidpunct}
{\mcitedefaultendpunct}{\mcitedefaultseppunct}\relax
\EndOfBibitem
\bibitem[T{\"o}rm{\"a} and Barnes(2015)]{torma_strong_2015}
P.~T{\"o}rm{\"a} and W.~L. Barnes, \emph{Reports on Progress in Physics}, 2015,
  \textbf{78}, 013901\relax
\mciteBstWouldAddEndPuncttrue
\mciteSetBstMidEndSepPunct{\mcitedefaultmidpunct}
{\mcitedefaultendpunct}{\mcitedefaultseppunct}\relax
\EndOfBibitem
\bibitem[Vasa and Lienau(2018)]{vasa2018}
P.~Vasa and C.~Lienau, \emph{ACS Photonics}, 2018, \textbf{5}, 2--23\relax
\mciteBstWouldAddEndPuncttrue
\mciteSetBstMidEndSepPunct{\mcitedefaultmidpunct}
{\mcitedefaultendpunct}{\mcitedefaultseppunct}\relax
\EndOfBibitem
\bibitem[Houdr{\'e} \emph{et~al.}(1996)Houdr{\'e}, Stanley, and
  Ilegems]{houdre_vacuum-field_1996}
R.~Houdr{\'e}, R.~P. Stanley and M.~Ilegems, \emph{Physical Review A}, 1996,
  \textbf{53}, 2711--2715\relax
\mciteBstWouldAddEndPuncttrue
\mciteSetBstMidEndSepPunct{\mcitedefaultmidpunct}
{\mcitedefaultendpunct}{\mcitedefaultseppunct}\relax
\EndOfBibitem
\bibitem[Agranovich \emph{et~al.}(2003)Agranovich, Litinskaia, and
  Lidzey]{agranovich2003}
V.~M. Agranovich, M.~Litinskaia and D.~G. Lidzey, \emph{Physical Review B},
  2003, \textbf{67}, 085311\relax
\mciteBstWouldAddEndPuncttrue
\mciteSetBstMidEndSepPunct{\mcitedefaultmidpunct}
{\mcitedefaultendpunct}{\mcitedefaultseppunct}\relax
\EndOfBibitem
\bibitem[Gonzalez-Ballestero \emph{et~al.}(2016)Gonzalez-Ballestero, Feist,
  Gonzalo~Bad{\'\i}a, Moreno, and Garcia-Vidal]{gonzalez-ballestero2016}
C.~Gonzalez-Ballestero, J.~Feist, E.~Gonzalo~Bad{\'\i}a, E.~Moreno and F.~J.
  Garcia-Vidal, \emph{Physical Review Letters}, 2016, \textbf{117},
  156402\relax
\mciteBstWouldAddEndPuncttrue
\mciteSetBstMidEndSepPunct{\mcitedefaultmidpunct}
{\mcitedefaultendpunct}{\mcitedefaultseppunct}\relax
\EndOfBibitem
\bibitem[Meschede \emph{et~al.}(1985)Meschede, Walther, and
  M{\"u}ller]{meschede1985}
D.~Meschede, H.~Walther and G.~M{\"u}ller, \emph{Physical Review Letters},
  1985, \textbf{54}, 551--554\relax
\mciteBstWouldAddEndPuncttrue
\mciteSetBstMidEndSepPunct{\mcitedefaultmidpunct}
{\mcitedefaultendpunct}{\mcitedefaultseppunct}\relax
\EndOfBibitem
\bibitem[Raizen \emph{et~al.}(1989)Raizen, Thompson, Brecha, Kimble, and
  Carmichael]{raizen1989}
M.~G. Raizen, R.~J. Thompson, R.~J. Brecha, H.~J. Kimble and H.~J. Carmichael,
  \emph{Physical Review Letters}, 1989, \textbf{63}, 240--243\relax
\mciteBstWouldAddEndPuncttrue
\mciteSetBstMidEndSepPunct{\mcitedefaultmidpunct}
{\mcitedefaultendpunct}{\mcitedefaultseppunct}\relax
\EndOfBibitem
\bibitem[Weisbuch \emph{et~al.}(1992)Weisbuch, Nishioka, Ishikawa, and
  Arakawa]{weisbuch_observation_1992}
C.~Weisbuch, M.~Nishioka, A.~Ishikawa and Y.~Arakawa, \emph{Physical Review
  Letters}, 1992, \textbf{69}, 3314--3317\relax
\mciteBstWouldAddEndPuncttrue
\mciteSetBstMidEndSepPunct{\mcitedefaultmidpunct}
{\mcitedefaultendpunct}{\mcitedefaultseppunct}\relax
\EndOfBibitem
\bibitem[Houdr{\'e} \emph{et~al.}(1993)Houdr{\'e}, Stanley, Oesterle, Ilegems,
  and Weisbuch]{houdre_room_1993}
R.~Houdr{\'e}, R.~P. Stanley, U.~Oesterle, M.~Ilegems and C.~Weisbuch, \emph{Le
  Journal de Physique IV}, 1993, \textbf{03}, C5--51--C5--58\relax
\mciteBstWouldAddEndPuncttrue
\mciteSetBstMidEndSepPunct{\mcitedefaultmidpunct}
{\mcitedefaultendpunct}{\mcitedefaultseppunct}\relax
\EndOfBibitem
\bibitem[Lidzey \emph{et~al.}(1998)Lidzey, Bradley, Skolnick, Virgili, Walker,
  and Whittaker]{lidzey1998}
D.~G. Lidzey, D.~D.~C. Bradley, M.~S. Skolnick, T.~Virgili, S.~Walker and D.~M.
  Whittaker, \emph{Nature}, 1998, \textbf{395}, 53\relax
\mciteBstWouldAddEndPuncttrue
\mciteSetBstMidEndSepPunct{\mcitedefaultmidpunct}
{\mcitedefaultendpunct}{\mcitedefaultseppunct}\relax
\EndOfBibitem
\bibitem[Schouwink \emph{et~al.}(2001)Schouwink, Berlepsch, D{\"a}hne, and
  Mahrt]{schouwink2001}
P.~Schouwink, H.~V. Berlepsch, L.~D{\"a}hne and R.~F. Mahrt, \emph{Chemical
  Physics Letters}, 2001, \textbf{344}, 352--356\relax
\mciteBstWouldAddEndPuncttrue
\mciteSetBstMidEndSepPunct{\mcitedefaultmidpunct}
{\mcitedefaultendpunct}{\mcitedefaultseppunct}\relax
\EndOfBibitem
\bibitem[Holmes and Forrest(2004)]{holmes2004a}
R.~J. Holmes and S.~R. Forrest, \emph{Physical Review Letters}, 2004,
  \textbf{93}, 186404\relax
\mciteBstWouldAddEndPuncttrue
\mciteSetBstMidEndSepPunct{\mcitedefaultmidpunct}
{\mcitedefaultendpunct}{\mcitedefaultseppunct}\relax
\EndOfBibitem
\bibitem[Dintinger \emph{et~al.}(2005)Dintinger, Klein, Bustos, Barnes, and
  Ebbesen]{dintinger_strong_2005}
J.~Dintinger, S.~Klein, F.~Bustos, W.~L. Barnes and T.~W. Ebbesen,
  \emph{Physical Review B}, 2005, \textbf{71}, 035424\relax
\mciteBstWouldAddEndPuncttrue
\mciteSetBstMidEndSepPunct{\mcitedefaultmidpunct}
{\mcitedefaultendpunct}{\mcitedefaultseppunct}\relax
\EndOfBibitem
\bibitem[K{\'e}na-Cohen \emph{et~al.}(2008)K{\'e}na-Cohen, Davan{\c c}o, and
  Forrest]{kena-cohen_strong_2008}
S.~K{\'e}na-Cohen, M.~Davan{\c c}o and S.~R. Forrest, \emph{Physical Review
  Letters}, 2008, \textbf{101}, 116401\relax
\mciteBstWouldAddEndPuncttrue
\mciteSetBstMidEndSepPunct{\mcitedefaultmidpunct}
{\mcitedefaultendpunct}{\mcitedefaultseppunct}\relax
\EndOfBibitem
\bibitem[K{\'e}na-Cohen and Forrest(2010)]{kena-cohen2010}
S.~K{\'e}na-Cohen and S.~R. Forrest, \emph{Nature Photonics}, 2010, \textbf{4},
  371--375\relax
\mciteBstWouldAddEndPuncttrue
\mciteSetBstMidEndSepPunct{\mcitedefaultmidpunct}
{\mcitedefaultendpunct}{\mcitedefaultseppunct}\relax
\EndOfBibitem
\bibitem[Virgili \emph{et~al.}(2011)Virgili, Coles, Adawi, Clark, Michetti,
  Rajendran, Brida, Polli, Cerullo, and Lidzey]{virgili2011}
T.~Virgili, D.~Coles, A.~M. Adawi, C.~Clark, P.~Michetti, S.~K. Rajendran,
  D.~Brida, D.~Polli, G.~Cerullo and D.~G. Lidzey, \emph{Physical Review B},
  2011, \textbf{83}, 245309\relax
\mciteBstWouldAddEndPuncttrue
\mciteSetBstMidEndSepPunct{\mcitedefaultmidpunct}
{\mcitedefaultendpunct}{\mcitedefaultseppunct}\relax
\EndOfBibitem
\bibitem[Schwartz \emph{et~al.}(2011)Schwartz, Hutchison, Genet, and
  Ebbesen]{schwartz2011}
T.~Schwartz, J.~A. Hutchison, C.~Genet and T.~W. Ebbesen, \emph{Physical Review
  Letters}, 2011, \textbf{106}, 196405\relax
\mciteBstWouldAddEndPuncttrue
\mciteSetBstMidEndSepPunct{\mcitedefaultmidpunct}
{\mcitedefaultendpunct}{\mcitedefaultseppunct}\relax
\EndOfBibitem
\bibitem[Aberra~Guebrou \emph{et~al.}(2012)Aberra~Guebrou, Symonds, Homeyer,
  Plenet, Gartstein, Agranovich, and Bellessa]{aberra_guebrou_coherent_2012}
S.~Aberra~Guebrou, C.~Symonds, E.~Homeyer, J.~C. Plenet, Y.~N. Gartstein, V.~M.
  Agranovich and J.~Bellessa, \emph{Physical Review Letters}, 2012,
  \textbf{108}, 066401\relax
\mciteBstWouldAddEndPuncttrue
\mciteSetBstMidEndSepPunct{\mcitedefaultmidpunct}
{\mcitedefaultendpunct}{\mcitedefaultseppunct}\relax
\EndOfBibitem
\bibitem[Hutchison \emph{et~al.}(2012)Hutchison, Schwartz, Genet, Devaux, and
  Ebbesen]{hutchison_modifying_2012}
J.~A. Hutchison, T.~Schwartz, C.~Genet, E.~Devaux and T.~W. Ebbesen,
  \emph{Angewandte Chemie International Edition}, 2012, \textbf{51},
  1592--1596\relax
\mciteBstWouldAddEndPuncttrue
\mciteSetBstMidEndSepPunct{\mcitedefaultmidpunct}
{\mcitedefaultendpunct}{\mcitedefaultseppunct}\relax
\EndOfBibitem
\bibitem[Schwartz \emph{et~al.}(2013)Schwartz, Hutchison, L{\'e}onard, Genet,
  Haacke, and Ebbesen]{schwartz_polariton_2013}
T.~Schwartz, J.~A. Hutchison, J.~L{\'e}onard, C.~Genet, S.~Haacke and T.~W.
  Ebbesen, \emph{ChemPhysChem}, 2013, \textbf{14}, 125--131\relax
\mciteBstWouldAddEndPuncttrue
\mciteSetBstMidEndSepPunct{\mcitedefaultmidpunct}
{\mcitedefaultendpunct}{\mcitedefaultseppunct}\relax
\EndOfBibitem
\bibitem[Simpkins \emph{et~al.}(2015)Simpkins, Fears, Dressick, Spann,
  Dunkelberger, and Owrutsky]{simpkins_spanning_2015}
B.~S. Simpkins, K.~P. Fears, W.~J. Dressick, B.~T. Spann, A.~D. Dunkelberger
  and J.~C. Owrutsky, \emph{ACS Photonics}, 2015, \textbf{2}, 1460--1467\relax
\mciteBstWouldAddEndPuncttrue
\mciteSetBstMidEndSepPunct{\mcitedefaultmidpunct}
{\mcitedefaultendpunct}{\mcitedefaultseppunct}\relax
\EndOfBibitem
\bibitem[Thomas \emph{et~al.}(2016)Thomas, George, Shalabney, Dryzhakov, Varma,
  Moran, Chervy, Zhong, Devaux, Genet, Hutchison, and
  Ebbesen]{thomas_ground-state_2016}
A.~Thomas, J.~George, A.~Shalabney, M.~Dryzhakov, S.~J. Varma, J.~Moran,
  T.~Chervy, X.~Zhong, E.~Devaux, C.~Genet, J.~A. Hutchison and T.~W. Ebbesen,
  \emph{Angewandte Chemie International Edition}, 2016, \textbf{55},
  11462--11466\relax
\mciteBstWouldAddEndPuncttrue
\mciteSetBstMidEndSepPunct{\mcitedefaultmidpunct}
{\mcitedefaultendpunct}{\mcitedefaultseppunct}\relax
\EndOfBibitem
\bibitem[Zhong \emph{et~al.}(2017)Zhong, Chervy, Zhang, Thomas, George, Genet,
  Hutchison, and Ebbesen]{zhong2017}
X.~Zhong, T.~Chervy, L.~Zhang, A.~Thomas, J.~George, C.~Genet, J.~A. Hutchison
  and T.~W. Ebbesen, \emph{Angewandte Chemie International Edition}, 2017,
  \textbf{56}, 9034--9038\relax
\mciteBstWouldAddEndPuncttrue
\mciteSetBstMidEndSepPunct{\mcitedefaultmidpunct}
{\mcitedefaultendpunct}{\mcitedefaultseppunct}\relax
\EndOfBibitem
\bibitem[Chikkaraddy \emph{et~al.}(2016)Chikkaraddy, {de Nijs}, Benz, Barrow,
  Scherman, Rosta, Demetriadou, Fox, Hess, and Baumberg]{chikkaraddy2016}
R.~Chikkaraddy, B.~{de Nijs}, F.~Benz, S.~J. Barrow, O.~A. Scherman, E.~Rosta,
  A.~Demetriadou, P.~Fox, O.~Hess and J.~J. Baumberg, \emph{Nature}, 2016,
  \textbf{535}, 127--130\relax
\mciteBstWouldAddEndPuncttrue
\mciteSetBstMidEndSepPunct{\mcitedefaultmidpunct}
{\mcitedefaultendpunct}{\mcitedefaultseppunct}\relax
\EndOfBibitem
\bibitem[Melnikau \emph{et~al.}(2017)Melnikau, Govyadinov,
  S{\'a}nchez-Iglesias, Grzelczak, Liz-Marz{\'a}n, and Rakovich]{melnikau2017}
D.~Melnikau, A.~A. Govyadinov, A.~S{\'a}nchez-Iglesias, M.~Grzelczak, L.~M.
  Liz-Marz{\'a}n and Y.~P. Rakovich, \emph{Nano Letters}, 2017, \textbf{17},
  1808--1813\relax
\mciteBstWouldAddEndPuncttrue
\mciteSetBstMidEndSepPunct{\mcitedefaultmidpunct}
{\mcitedefaultendpunct}{\mcitedefaultseppunct}\relax
\EndOfBibitem
\bibitem[Baieva \emph{et~al.}(2017)Baieva, Hakamaa, Groenhof, Heikkil{\"a}, and
  Toppari]{baieva_dynamics_2017}
S.~Baieva, O.~Hakamaa, G.~Groenhof, T.~T. Heikkil{\"a} and J.~J. Toppari,
  \emph{ACS Photonics}, 2017, \textbf{4}, 28--37\relax
\mciteBstWouldAddEndPuncttrue
\mciteSetBstMidEndSepPunct{\mcitedefaultmidpunct}
{\mcitedefaultendpunct}{\mcitedefaultseppunct}\relax
\EndOfBibitem
\bibitem[Crum \emph{et~al.}(2018)Crum, Casey, and Sparks]{crum2018}
V.~F. Crum, S.~R. Casey and J.~R. Sparks, \emph{Physical Chemistry Chemical
  Physics}, 2018, \textbf{20}, 850--857\relax
\mciteBstWouldAddEndPuncttrue
\mciteSetBstMidEndSepPunct{\mcitedefaultmidpunct}
{\mcitedefaultendpunct}{\mcitedefaultseppunct}\relax
\EndOfBibitem
\bibitem[Rozenman \emph{et~al.}(2018)Rozenman, Akulov, Golombek, and
  Schwartz]{rozenman2018}
G.~G. Rozenman, K.~Akulov, A.~Golombek and T.~Schwartz, \emph{ACS Photonics},
  2018, \textbf{5}, 105--110\relax
\mciteBstWouldAddEndPuncttrue
\mciteSetBstMidEndSepPunct{\mcitedefaultmidpunct}
{\mcitedefaultendpunct}{\mcitedefaultseppunct}\relax
\EndOfBibitem
\bibitem[Dunkelberger \emph{et~al.}(2018)Dunkelberger, Davidson~II, Ahn,
  Simpkins, and Owrutsky]{dunkelberger2018b}
A.~D. Dunkelberger, R.~B. Davidson~II, W.~Ahn, B.~S. Simpkins and J.~C.
  Owrutsky, \emph{The Journal of Physical Chemistry A}, 2018\relax
\mciteBstWouldAddEndPuncttrue
\mciteSetBstMidEndSepPunct{\mcitedefaultmidpunct}
{\mcitedefaultendpunct}{\mcitedefaultseppunct}\relax
\EndOfBibitem
\bibitem[Cheng \emph{et~al.}(2018)Cheng, Dhanker, Gray, Mukhopadhyay, Kennehan,
  Asbury, Sokolov, and Giebink]{cheng2018}
C.-Y. Cheng, R.~Dhanker, C.~L. Gray, S.~Mukhopadhyay, E.~R. Kennehan, J.~B.
  Asbury, A.~Sokolov and N.~C. Giebink, \emph{Physical Review Letters}, 2018,
  \textbf{120}, 017402\relax
\mciteBstWouldAddEndPuncttrue
\mciteSetBstMidEndSepPunct{\mcitedefaultmidpunct}
{\mcitedefaultendpunct}{\mcitedefaultseppunct}\relax
\EndOfBibitem
\bibitem[{\'C}wik \emph{et~al.}(2014){\'C}wik, Reja, Littlewood, and
  Keeling]{cwik2014}
J.~A. {\'C}wik, S.~Reja, P.~B. Littlewood and J.~Keeling, \emph{EPL
  (Europhysics Letters)}, 2014, \textbf{105}, 47009\relax
\mciteBstWouldAddEndPuncttrue
\mciteSetBstMidEndSepPunct{\mcitedefaultmidpunct}
{\mcitedefaultendpunct}{\mcitedefaultseppunct}\relax
\EndOfBibitem
\bibitem[del Pino \emph{et~al.}(2015)del Pino, Feist, and
  Garcia-Vidal]{pino2015}
J.~del Pino, J.~Feist and F.~J. Garcia-Vidal, \emph{New Journal of Physics},
  2015, \textbf{17}, 053040\relax
\mciteBstWouldAddEndPuncttrue
\mciteSetBstMidEndSepPunct{\mcitedefaultmidpunct}
{\mcitedefaultendpunct}{\mcitedefaultseppunct}\relax
\EndOfBibitem
\bibitem[Herrera and Spano(2016)]{herrera2016}
F.~Herrera and F.~C. Spano, \emph{Physical Review Letters}, 2016, \textbf{116},
  238301\relax
\mciteBstWouldAddEndPuncttrue
\mciteSetBstMidEndSepPunct{\mcitedefaultmidpunct}
{\mcitedefaultendpunct}{\mcitedefaultseppunct}\relax
\EndOfBibitem
\bibitem[Galego \emph{et~al.}(2016)Galego, Garcia-Vidal, and Feist]{galego2016}
J.~Galego, F.~J. Garcia-Vidal and J.~Feist, \emph{Nature Communications}, 2016,
  \textbf{7}:13841\relax
\mciteBstWouldAddEndPuncttrue
\mciteSetBstMidEndSepPunct{\mcitedefaultmidpunct}
{\mcitedefaultendpunct}{\mcitedefaultseppunct}\relax
\EndOfBibitem
\bibitem[Kowalewski \emph{et~al.}(2016)Kowalewski, Bennett, and
  Mukamel]{kowalewski2016}
M.~Kowalewski, K.~Bennett and S.~Mukamel, \emph{The Journal of Chemical
  Physics}, 2016, \textbf{144}, 054309\relax
\mciteBstWouldAddEndPuncttrue
\mciteSetBstMidEndSepPunct{\mcitedefaultmidpunct}
{\mcitedefaultendpunct}{\mcitedefaultseppunct}\relax
\EndOfBibitem
\bibitem[Bennett \emph{et~al.}(2016)Bennett, Kowalewski, and
  Mukamel]{bennett2016a}
K.~Bennett, M.~Kowalewski and S.~Mukamel, \emph{Faraday Discussions}, 2016,
  \textbf{194}, 259--282\relax
\mciteBstWouldAddEndPuncttrue
\mciteSetBstMidEndSepPunct{\mcitedefaultmidpunct}
{\mcitedefaultendpunct}{\mcitedefaultseppunct}\relax
\EndOfBibitem
\bibitem[Wu \emph{et~al.}(2016)Wu, Feist, and Garcia-Vidal]{wu2016}
N.~Wu, J.~Feist and F.~J. Garcia-Vidal, \emph{Physical Review B}, 2016,
  \textbf{94}, 195409\relax
\mciteBstWouldAddEndPuncttrue
\mciteSetBstMidEndSepPunct{\mcitedefaultmidpunct}
{\mcitedefaultendpunct}{\mcitedefaultseppunct}\relax
\EndOfBibitem
\bibitem[Herrera and Spano(2017)]{herrera_dark_2017}
F.~Herrera and F.~C. Spano, \emph{Physical Review Letters}, 2017, \textbf{118},
  223601\relax
\mciteBstWouldAddEndPuncttrue
\mciteSetBstMidEndSepPunct{\mcitedefaultmidpunct}
{\mcitedefaultendpunct}{\mcitedefaultseppunct}\relax
\EndOfBibitem
\bibitem[Zhang and Mukamel(2017)]{zhang2017}
Z.~Zhang and S.~Mukamel, \emph{Chemical Physics Letters}, 2017, \textbf{683},
  653--657\relax
\mciteBstWouldAddEndPuncttrue
\mciteSetBstMidEndSepPunct{\mcitedefaultmidpunct}
{\mcitedefaultendpunct}{\mcitedefaultseppunct}\relax
\EndOfBibitem
\bibitem[Feist \emph{et~al.}(2017)Feist, Galego, and Garcia-Vidal]{feist2017}
J.~Feist, J.~Galego and F.~J. Garcia-Vidal, \emph{ACS Photonics}, 2017\relax
\mciteBstWouldAddEndPuncttrue
\mciteSetBstMidEndSepPunct{\mcitedefaultmidpunct}
{\mcitedefaultendpunct}{\mcitedefaultseppunct}\relax
\EndOfBibitem
\bibitem[Dimitrov \emph{et~al.}(2017)Dimitrov, Flick, Ruggenthaler, and
  Rubio]{dimitrov2017}
T.~Dimitrov, J.~Flick, M.~Ruggenthaler and A.~Rubio, \emph{New Journal of
  Physics}, 2017, \textbf{19}, 113036\relax
\mciteBstWouldAddEndPuncttrue
\mciteSetBstMidEndSepPunct{\mcitedefaultmidpunct}
{\mcitedefaultendpunct}{\mcitedefaultseppunct}\relax
\EndOfBibitem
\bibitem[Flick \emph{et~al.}(2017)Flick, Appel, Ruggenthaler, and
  Rubio]{flick_cavity_2017}
J.~Flick, H.~Appel, M.~Ruggenthaler and A.~Rubio, \emph{Journal of Chemical
  Theory and Computation}, 2017, \textbf{13}, 1616--1625\relax
\mciteBstWouldAddEndPuncttrue
\mciteSetBstMidEndSepPunct{\mcitedefaultmidpunct}
{\mcitedefaultendpunct}{\mcitedefaultseppunct}\relax
\EndOfBibitem
\bibitem[Zeb \emph{et~al.}(2018)Zeb, Kirton, and Keeling]{zeb2018}
M.~A. Zeb, P.~G. Kirton and J.~Keeling, \emph{ACS Photonics}, 2018, \textbf{5},
  249--257\relax
\mciteBstWouldAddEndPuncttrue
\mciteSetBstMidEndSepPunct{\mcitedefaultmidpunct}
{\mcitedefaultendpunct}{\mcitedefaultseppunct}\relax
\EndOfBibitem
\bibitem[Yuen-Zhou \emph{et~al.}(2017)Yuen-Zhou, Saikin, and
  Menon]{yuen-zhou2017}
J.~Yuen-Zhou, S.~K. Saikin and V.~Menon, \emph{arXiv:1711.11213 [cond-mat]},
  2017\relax
\mciteBstWouldAddEndPuncttrue
\mciteSetBstMidEndSepPunct{\mcitedefaultmidpunct}
{\mcitedefaultendpunct}{\mcitedefaultseppunct}\relax
\EndOfBibitem
\bibitem[Mart{\'\i}nez-Mart{\'\i}nez
  \emph{et~al.}(2018)Mart{\'\i}nez-Mart{\'\i}nez, Ribeiro,
  Campos-Gonz{\'a}lez-Angulo, and Yuen-Zhou]{martinez-martinez2018}
L.~A. Mart{\'\i}nez-Mart{\'\i}nez, R.~F. Ribeiro, J.~Campos-Gonz{\'a}lez-Angulo
  and J.~Yuen-Zhou, \emph{ACS Photonics}, 2018, \textbf{5}, 167--176\relax
\mciteBstWouldAddEndPuncttrue
\mciteSetBstMidEndSepPunct{\mcitedefaultmidpunct}
{\mcitedefaultendpunct}{\mcitedefaultseppunct}\relax
\EndOfBibitem
\bibitem[Du \emph{et~al.}(2017)Du, Mart{\'\i}nez-Mart{\'\i}nez, Ribeiro, Hu,
  Menon, and Yuen-Zhou]{du2017}
M.~Du, L.~A. Mart{\'\i}nez-Mart{\'\i}nez, R.~F. Ribeiro, Z.~Hu, V.~M. Menon and
  J.~Yuen-Zhou, \emph{arXiv:1711.11576 [cond-mat, physics:quant-ph]},
  2017\relax
\mciteBstWouldAddEndPuncttrue
\mciteSetBstMidEndSepPunct{\mcitedefaultmidpunct}
{\mcitedefaultendpunct}{\mcitedefaultseppunct}\relax
\EndOfBibitem
\bibitem[Martinez-Martinez \emph{et~al.}(2017)Martinez-Martinez, Du, Ribeiro,
  K{\'e}na-Cohen, and Yuen-Zhou]{martinez-martinez2017b}
L.~A. Martinez-Martinez, M.~Du, R.~F. Ribeiro, S.~K{\'e}na-Cohen and
  J.~Yuen-Zhou, \emph{arXiv:1711.11264 [physics, physics:quant-ph]}, 2017\relax
\mciteBstWouldAddEndPuncttrue
\mciteSetBstMidEndSepPunct{\mcitedefaultmidpunct}
{\mcitedefaultendpunct}{\mcitedefaultseppunct}\relax
\EndOfBibitem
\bibitem[Ribeiro \emph{et~al.}(2017)Ribeiro, Dunkelberger, Xiang, Xiong,
  Simpkins, Owrutsky, and Yuen-Zhou]{ribeiro2017c}
R.~F. Ribeiro, A.~D. Dunkelberger, B.~Xiang, W.~Xiong, B.~S. Simpkins, J.~C.
  Owrutsky and J.~Yuen-Zhou, \emph{arXiv:1711.11242 [physics,
  physics:quant-ph]}, 2017\relax
\mciteBstWouldAddEndPuncttrue
\mciteSetBstMidEndSepPunct{\mcitedefaultmidpunct}
{\mcitedefaultendpunct}{\mcitedefaultseppunct}\relax
\EndOfBibitem
\bibitem[Carusotto and Ciuti(2013)]{carusotto_quantum_2013}
I.~Carusotto and C.~Ciuti, \emph{Reviews of Modern Physics}, 2013, \textbf{85},
  299--366\relax
\mciteBstWouldAddEndPuncttrue
\mciteSetBstMidEndSepPunct{\mcitedefaultmidpunct}
{\mcitedefaultendpunct}{\mcitedefaultseppunct}\relax
\EndOfBibitem
\bibitem[Ruggenthaler \emph{et~al.}(2014)Ruggenthaler, Flick, Pellegrini,
  Appel, Tokatly, and Rubio]{ruggenthaler2014}
M.~Ruggenthaler, J.~Flick, C.~Pellegrini, H.~Appel, I.~V. Tokatly and A.~Rubio,
  \emph{Physical Review A}, 2014, \textbf{90}, 012508\relax
\mciteBstWouldAddEndPuncttrue
\mciteSetBstMidEndSepPunct{\mcitedefaultmidpunct}
{\mcitedefaultendpunct}{\mcitedefaultseppunct}\relax
\EndOfBibitem
\bibitem[Holmes and Forrest(2007)]{holmes2007}
R.~J. Holmes and S.~R. Forrest, \emph{Organic Electronics}, 2007, \textbf{8},
  77--93\relax
\mciteBstWouldAddEndPuncttrue
\mciteSetBstMidEndSepPunct{\mcitedefaultmidpunct}
{\mcitedefaultendpunct}{\mcitedefaultseppunct}\relax
\EndOfBibitem
\bibitem[Agranovich \emph{et~al.}(2011)Agranovich, Gartstein, and
  Litinskaya]{agranovich2011}
V.~M. Agranovich, Y.~N. Gartstein and M.~Litinskaya, \emph{Chemical Reviews},
  2011, \textbf{111}, 5179--5214\relax
\mciteBstWouldAddEndPuncttrue
\mciteSetBstMidEndSepPunct{\mcitedefaultmidpunct}
{\mcitedefaultendpunct}{\mcitedefaultseppunct}\relax
\EndOfBibitem
\bibitem[K{\'e}na-Cohen and Forrest(2012)]{kena-cohen2012}
S.~K{\'e}na-Cohen and S.~R. Forrest, \emph{Exciton {{Polaritons}} in
  {{Microcavities}}}, {Springer, Berlin, Heidelberg}, 2012, pp. 349--375\relax
\mciteBstWouldAddEndPuncttrue
\mciteSetBstMidEndSepPunct{\mcitedefaultmidpunct}
{\mcitedefaultendpunct}{\mcitedefaultseppunct}\relax
\EndOfBibitem
\bibitem[Sukharev and Nitzan(2017)]{sukharev_optics_2017}
M.~Sukharev and A.~Nitzan, \emph{Journal of Physics: Condensed Matter}, 2017,
  \textbf{29}, 443003\relax
\mciteBstWouldAddEndPuncttrue
\mciteSetBstMidEndSepPunct{\mcitedefaultmidpunct}
{\mcitedefaultendpunct}{\mcitedefaultseppunct}\relax
\EndOfBibitem
\bibitem[Herrera and Spano(2018)]{herrera2018}
F.~Herrera and F.~C. Spano, \emph{ACS Photonics}, 2018, \textbf{5},
  65--79\relax
\mciteBstWouldAddEndPuncttrue
\mciteSetBstMidEndSepPunct{\mcitedefaultmidpunct}
{\mcitedefaultendpunct}{\mcitedefaultseppunct}\relax
\EndOfBibitem
\bibitem[S.~Dovzhenko \emph{et~al.}(2018)S.~Dovzhenko, V.~Ryabchuk,
  P.~Rakovich, and R.~Nabiev]{s.dovzhenko2018}
D.~S.~Dovzhenko, S.~V.~Ryabchuk, Y.~P.~Rakovich and I.~R.~Nabiev,
  \emph{Nanoscale}, 2018, \textbf{10}, 3589--3605\relax
\mciteBstWouldAddEndPuncttrue
\mciteSetBstMidEndSepPunct{\mcitedefaultmidpunct}
{\mcitedefaultendpunct}{\mcitedefaultseppunct}\relax
\EndOfBibitem
\bibitem[Steck(2007)]{steck2007quantum}
D.~A. Steck, \emph{Oregon Center for Optics and Department of Physics,
  University of Oregon}, 2007\relax
\mciteBstWouldAddEndPuncttrue
\mciteSetBstMidEndSepPunct{\mcitedefaultmidpunct}
{\mcitedefaultendpunct}{\mcitedefaultseppunct}\relax
\EndOfBibitem
\bibitem[Coles and Lidzey(2014)]{coles2014}
D.~M. Coles and D.~G. Lidzey, \emph{Applied Physics Letters}, 2014,
  \textbf{104}, 191108\relax
\mciteBstWouldAddEndPuncttrue
\mciteSetBstMidEndSepPunct{\mcitedefaultmidpunct}
{\mcitedefaultendpunct}{\mcitedefaultseppunct}\relax
\EndOfBibitem
\bibitem[Daskalakis \emph{et~al.}(2017)Daskalakis, Maier, and
  K{\'e}na-Cohen]{Daskalakis2017}
K.~S. Daskalakis, S.~A. Maier and S.~K{\'e}na-Cohen, \emph{Quantum
  {{Plasmonics}}}, {Springer International Publishing}, Cham, 2017, pp.
  151--163\relax
\mciteBstWouldAddEndPuncttrue
\mciteSetBstMidEndSepPunct{\mcitedefaultmidpunct}
{\mcitedefaultendpunct}{\mcitedefaultseppunct}\relax
\EndOfBibitem
\bibitem[Jaynes and Cummings(1963)]{jaynes_comparison_1963}
E.~T. Jaynes and F.~W. Cummings, \emph{Proceedings of the IEEE}, 1963,
  \textbf{51}, 89--109\relax
\mciteBstWouldAddEndPuncttrue
\mciteSetBstMidEndSepPunct{\mcitedefaultmidpunct}
{\mcitedefaultendpunct}{\mcitedefaultseppunct}\relax
\EndOfBibitem
\bibitem[Ujihara \emph{et~al.}(1991)Ujihara, Nakamura, and Manba]{ujihara1991}
K.~Ujihara, A.~Nakamura and O.~Manba, \emph{Japanese Journal of Applied
  Physics}, 1991, \textbf{30}, 3388\relax
\mciteBstWouldAddEndPuncttrue
\mciteSetBstMidEndSepPunct{\mcitedefaultmidpunct}
{\mcitedefaultendpunct}{\mcitedefaultseppunct}\relax
\EndOfBibitem
\bibitem[Rabi(1937)]{rabi1937}
I.~I. Rabi, \emph{Physical Review}, 1937, \textbf{51}, 652--654\relax
\mciteBstWouldAddEndPuncttrue
\mciteSetBstMidEndSepPunct{\mcitedefaultmidpunct}
{\mcitedefaultendpunct}{\mcitedefaultseppunct}\relax
\EndOfBibitem
\bibitem[Agarwal(1984)]{agarwal1984}
G.~S. Agarwal, \emph{Physical Review Letters}, 1984, \textbf{53},
  1732--1734\relax
\mciteBstWouldAddEndPuncttrue
\mciteSetBstMidEndSepPunct{\mcitedefaultmidpunct}
{\mcitedefaultendpunct}{\mcitedefaultseppunct}\relax
\EndOfBibitem
\bibitem[Birnbaum \emph{et~al.}(2005)Birnbaum, Boca, Miller, Boozer, Northup,
  and Kimble]{birnbaum2005}
K.~M. Birnbaum, A.~Boca, R.~Miller, A.~D. Boozer, T.~E. Northup and H.~J.
  Kimble, \emph{Nature}, 2005, \textbf{436}, 87--90\relax
\mciteBstWouldAddEndPuncttrue
\mciteSetBstMidEndSepPunct{\mcitedefaultmidpunct}
{\mcitedefaultendpunct}{\mcitedefaultseppunct}\relax
\EndOfBibitem
\bibitem[Benz \emph{et~al.}(2016)Benz, Schmidt, Dreismann, Chikkaraddy, Zhang,
  Demetriadou, Carnegie, Ohadi, de~Nijs, Esteban, Aizpurua, and
  Baumberg]{benz_single-molecule_2016}
F.~Benz, M.~K. Schmidt, A.~Dreismann, R.~Chikkaraddy, Y.~Zhang, A.~Demetriadou,
  C.~Carnegie, H.~Ohadi, B.~de~Nijs, R.~Esteban, J.~Aizpurua and J.~J.
  Baumberg, \emph{Science}, 2016, \textbf{354}, 726--729\relax
\mciteBstWouldAddEndPuncttrue
\mciteSetBstMidEndSepPunct{\mcitedefaultmidpunct}
{\mcitedefaultendpunct}{\mcitedefaultseppunct}\relax
\EndOfBibitem
\bibitem[Wang \emph{et~al.}(2017)Wang, Kelkar, Martin-Cano, Utikal,
  G{\"o}tzinger, and Sandoghdar]{wang2017}
D.~Wang, H.~Kelkar, D.~Martin-Cano, T.~Utikal, S.~G{\"o}tzinger and
  V.~Sandoghdar, \emph{Physical Review X}, 2017, \textbf{7}, 021014\relax
\mciteBstWouldAddEndPuncttrue
\mciteSetBstMidEndSepPunct{\mcitedefaultmidpunct}
{\mcitedefaultendpunct}{\mcitedefaultseppunct}\relax
\EndOfBibitem
\bibitem[Tavis and Cummings(1968)]{tavis_exact_1968}
M.~Tavis and F.~W. Cummings, \emph{Physical Review}, 1968, \textbf{170},
  379--384\relax
\mciteBstWouldAddEndPuncttrue
\mciteSetBstMidEndSepPunct{\mcitedefaultmidpunct}
{\mcitedefaultendpunct}{\mcitedefaultseppunct}\relax
\EndOfBibitem
\bibitem[Tavis and Cummings(1969)]{tavis_approximate_1969}
M.~Tavis and F.~W. Cummings, \emph{Physical Review}, 1969, \textbf{188},
  692--695\relax
\mciteBstWouldAddEndPuncttrue
\mciteSetBstMidEndSepPunct{\mcitedefaultmidpunct}
{\mcitedefaultendpunct}{\mcitedefaultseppunct}\relax
\EndOfBibitem
\bibitem[Dicke(1954)]{dicke_coherence_1954}
R.~H. Dicke, \emph{Physical Review}, 1954, \textbf{93}, 99--110\relax
\mciteBstWouldAddEndPuncttrue
\mciteSetBstMidEndSepPunct{\mcitedefaultmidpunct}
{\mcitedefaultendpunct}{\mcitedefaultseppunct}\relax
\EndOfBibitem
\bibitem[Vetter \emph{et~al.}(2016)Vetter, Wang, Wang, and
  Scully]{vetter2016single}
P.~A. Vetter, L.~Wang, D.-W. Wang and M.~O. Scully, \emph{Physica Scripta},
  2016, \textbf{91}, 023007\relax
\mciteBstWouldAddEndPuncttrue
\mciteSetBstMidEndSepPunct{\mcitedefaultmidpunct}
{\mcitedefaultendpunct}{\mcitedefaultseppunct}\relax
\EndOfBibitem
\bibitem[Agarwal(1998)]{agarwal1998}
G.~S. Agarwal, \emph{Journal of Modern Optics}, 1998, \textbf{45},
  449--470\relax
\mciteBstWouldAddEndPuncttrue
\mciteSetBstMidEndSepPunct{\mcitedefaultmidpunct}
{\mcitedefaultendpunct}{\mcitedefaultseppunct}\relax
\EndOfBibitem
\bibitem[Manceau \emph{et~al.}(2017)Manceau, Biasiol, Tran, Carusotto, and
  Colombelli]{manceau2017}
J.-M. Manceau, G.~Biasiol, N.~L. Tran, I.~Carusotto and R.~Colombelli,
  \emph{Physical Review B}, 2017, \textbf{96}, 235301\relax
\mciteBstWouldAddEndPuncttrue
\mciteSetBstMidEndSepPunct{\mcitedefaultmidpunct}
{\mcitedefaultendpunct}{\mcitedefaultseppunct}\relax
\EndOfBibitem
\bibitem[Litinskaya and Reineker(2006)]{litinskaya_loss_2006}
M.~Litinskaya and P.~Reineker, \emph{Physical Review B}, 2006, \textbf{74},
  165320\relax
\mciteBstWouldAddEndPuncttrue
\mciteSetBstMidEndSepPunct{\mcitedefaultmidpunct}
{\mcitedefaultendpunct}{\mcitedefaultseppunct}\relax
\EndOfBibitem
\bibitem[Agranovich and Gartstein(2007)]{agranovich_nature_2007}
V.~M. Agranovich and Y.~N. Gartstein, \emph{Physical Review B}, 2007,
  \textbf{75}, 075302\relax
\mciteBstWouldAddEndPuncttrue
\mciteSetBstMidEndSepPunct{\mcitedefaultmidpunct}
{\mcitedefaultendpunct}{\mcitedefaultseppunct}\relax
\EndOfBibitem
\bibitem[Litinskaya(2008)]{litinskaya_propagation_2008}
M.~Litinskaya, \emph{Physics Letters A}, 2008, \textbf{372}, 3898--3903\relax
\mciteBstWouldAddEndPuncttrue
\mciteSetBstMidEndSepPunct{\mcitedefaultmidpunct}
{\mcitedefaultendpunct}{\mcitedefaultseppunct}\relax
\EndOfBibitem
\bibitem[Agranovich(2009)]{agranovich2009excitations}
V.~M. Agranovich, \emph{Excitations in Organic Solids}, {OUP Oxford}, 2009,
  vol. 142\relax
\mciteBstWouldAddEndPuncttrue
\mciteSetBstMidEndSepPunct{\mcitedefaultmidpunct}
{\mcitedefaultendpunct}{\mcitedefaultseppunct}\relax
\EndOfBibitem
\bibitem[Lidzey \emph{et~al.}(1999)Lidzey, Bradley, Virgili, Armitage,
  Skolnick, and Walker]{lidzey1999}
D.~G. Lidzey, D.~D.~C. Bradley, T.~Virgili, A.~Armitage, M.~S. Skolnick and
  S.~Walker, \emph{Physical Review Letters}, 1999, \textbf{82},
  3316--3319\relax
\mciteBstWouldAddEndPuncttrue
\mciteSetBstMidEndSepPunct{\mcitedefaultmidpunct}
{\mcitedefaultendpunct}{\mcitedefaultseppunct}\relax
\EndOfBibitem
\bibitem[Plumhof \emph{et~al.}(2014)Plumhof, St{\"o}ferle, Mai, Scherf, and
  Mahrt]{plumhof_room-temperature_2014}
J.~D. Plumhof, T.~St{\"o}ferle, L.~Mai, U.~Scherf and R.~F. Mahrt, \emph{Nature
  Materials}, 2014, \textbf{13}, 247--252\relax
\mciteBstWouldAddEndPuncttrue
\mciteSetBstMidEndSepPunct{\mcitedefaultmidpunct}
{\mcitedefaultendpunct}{\mcitedefaultseppunct}\relax
\EndOfBibitem
\bibitem[Orgiu \emph{et~al.}(2015)Orgiu, George, Hutchison, Devaux, Dayen,
  Doudin, Stellacci, Genet, Schachenmayer, Genes, Pupillo, Samor{\`\i}, and
  Ebbesen]{orgiu_conductivity_2015}
E.~Orgiu, J.~George, J.~A. Hutchison, E.~Devaux, J.~F. Dayen, B.~Doudin,
  F.~Stellacci, C.~Genet, J.~Schachenmayer, C.~Genes, G.~Pupillo,
  P.~Samor{\`\i} and T.~W. Ebbesen, \emph{Nature Materials}, 2015, \textbf{14},
  1123--1129\relax
\mciteBstWouldAddEndPuncttrue
\mciteSetBstMidEndSepPunct{\mcitedefaultmidpunct}
{\mcitedefaultendpunct}{\mcitedefaultseppunct}\relax
\EndOfBibitem
\bibitem[Coles \emph{et~al.}(2014)Coles, Somaschi, Michetti, Clark, Lagoudakis,
  Savvidis, and Lidzey]{coles2014b}
D.~M. Coles, N.~Somaschi, P.~Michetti, C.~Clark, P.~G. Lagoudakis, P.~G.
  Savvidis and D.~G. Lidzey, \emph{Nature Materials}, 2014, \textbf{13},
  712\relax
\mciteBstWouldAddEndPuncttrue
\mciteSetBstMidEndSepPunct{\mcitedefaultmidpunct}
{\mcitedefaultendpunct}{\mcitedefaultseppunct}\relax
\EndOfBibitem
\bibitem[Georgiou \emph{et~al.}(2018)Georgiou, Michetti, Gai, Cavazzini, Shen,
  and Lidzey]{georgiou2018}
K.~Georgiou, P.~Michetti, L.~Gai, M.~Cavazzini, Z.~Shen and D.~G. Lidzey,
  \emph{ACS Photonics}, 2018, \textbf{5}, 258--266\relax
\mciteBstWouldAddEndPuncttrue
\mciteSetBstMidEndSepPunct{\mcitedefaultmidpunct}
{\mcitedefaultendpunct}{\mcitedefaultseppunct}\relax
\EndOfBibitem
\bibitem[Hobson \emph{et~al.}(2002)Hobson, Barnes, Lidzey, Gehring, Whittaker,
  Skolnick, and Walker]{hobson2002}
P.~A. Hobson, W.~L. Barnes, D.~G. Lidzey, G.~A. Gehring, D.~M. Whittaker, M.~S.
  Skolnick and S.~Walker, \emph{Applied Physics Letters}, 2002, \textbf{81},
  3519--3521\relax
\mciteBstWouldAddEndPuncttrue
\mciteSetBstMidEndSepPunct{\mcitedefaultmidpunct}
{\mcitedefaultendpunct}{\mcitedefaultseppunct}\relax
\EndOfBibitem
\bibitem[Coles \emph{et~al.}(2011)Coles, Michetti, Clark, Tsoi, Adawi, Kim, and
  Lidzey]{coles2011a}
D.~M. Coles, P.~Michetti, C.~Clark, W.~C. Tsoi, A.~M. Adawi, J.-S. Kim and
  D.~G. Lidzey, \emph{Advanced Functional Materials}, 2011, \textbf{21},
  3691--3696\relax
\mciteBstWouldAddEndPuncttrue
\mciteSetBstMidEndSepPunct{\mcitedefaultmidpunct}
{\mcitedefaultendpunct}{\mcitedefaultseppunct}\relax
\EndOfBibitem
\bibitem[Michetti and La~Rocca(2008)]{michetti2008}
P.~Michetti and G.~C. La~Rocca, \emph{Physical Review B}, 2008, \textbf{77},
  195301\relax
\mciteBstWouldAddEndPuncttrue
\mciteSetBstMidEndSepPunct{\mcitedefaultmidpunct}
{\mcitedefaultendpunct}{\mcitedefaultseppunct}\relax
\EndOfBibitem
\bibitem[Taylor \emph{et~al.}(1984)Taylor, Erskine, and Tang]{taylor1984}
A.~J. Taylor, D.~J. Erskine and C.~L. Tang, \emph{Chemical Physics Letters},
  1984, \textbf{103}, 430--435\relax
\mciteBstWouldAddEndPuncttrue
\mciteSetBstMidEndSepPunct{\mcitedefaultmidpunct}
{\mcitedefaultendpunct}{\mcitedefaultseppunct}\relax
\EndOfBibitem
\bibitem[Reiser and Laubereau(1982)]{reiser1982}
D.~Reiser and A.~Laubereau, \emph{Optics Communications}, 1982, \textbf{42},
  329--334\relax
\mciteBstWouldAddEndPuncttrue
\mciteSetBstMidEndSepPunct{\mcitedefaultmidpunct}
{\mcitedefaultendpunct}{\mcitedefaultseppunct}\relax
\EndOfBibitem
\bibitem[Litinskaya \emph{et~al.}(2004)Litinskaya, Reineker, and
  Agranovich]{litinskaya2004}
M.~Litinskaya, P.~Reineker and V.~M. Agranovich, \emph{Journal of
  Luminescence}, 2004, \textbf{110}, 364--372\relax
\mciteBstWouldAddEndPuncttrue
\mciteSetBstMidEndSepPunct{\mcitedefaultmidpunct}
{\mcitedefaultendpunct}{\mcitedefaultseppunct}\relax
\EndOfBibitem
\bibitem[Coles \emph{et~al.}(2011)Coles, Michetti, Clark, Adawi, and
  Lidzey]{coles2011}
D.~M. Coles, P.~Michetti, C.~Clark, A.~M. Adawi and D.~G. Lidzey,
  \emph{Physical Review B}, 2011, \textbf{84}, 205214\relax
\mciteBstWouldAddEndPuncttrue
\mciteSetBstMidEndSepPunct{\mcitedefaultmidpunct}
{\mcitedefaultendpunct}{\mcitedefaultseppunct}\relax
\EndOfBibitem
\bibitem[Wang \emph{et~al.}(2014)Wang, Chervy, George, Hutchison, Genet, and
  Ebbesen]{wang2014}
S.~Wang, T.~Chervy, J.~George, J.~A. Hutchison, C.~Genet and T.~W. Ebbesen,
  \emph{The Journal of Physical Chemistry Letters}, 2014, \textbf{5},
  1433--1439\relax
\mciteBstWouldAddEndPuncttrue
\mciteSetBstMidEndSepPunct{\mcitedefaultmidpunct}
{\mcitedefaultendpunct}{\mcitedefaultseppunct}\relax
\EndOfBibitem
\bibitem[Furuki \emph{et~al.}(2001)Furuki, Tian, Sato, Pu, Kawashima, Tatsuura,
  and Wada]{furuki2001}
M.~Furuki, M.~Tian, Y.~Sato, L.~S. Pu, H.~Kawashima, S.~Tatsuura and O.~Wada,
  \emph{Applied Physics Letters}, 2001, \textbf{78}, 2634--2636\relax
\mciteBstWouldAddEndPuncttrue
\mciteSetBstMidEndSepPunct{\mcitedefaultmidpunct}
{\mcitedefaultendpunct}{\mcitedefaultseppunct}\relax
\EndOfBibitem
\bibitem[Song \emph{et~al.}(2004)Song, He, Nurmikko, Tischler, and
  Bulovic]{song2004}
J.-H. Song, Y.~He, A.~V. Nurmikko, J.~Tischler and V.~Bulovic, \emph{Physical
  Review B}, 2004, \textbf{69}, 235330\relax
\mciteBstWouldAddEndPuncttrue
\mciteSetBstMidEndSepPunct{\mcitedefaultmidpunct}
{\mcitedefaultendpunct}{\mcitedefaultseppunct}\relax
\EndOfBibitem
\bibitem[May and K{\"u}hn(2004)]{may2004charge}
V.~May and O.~K{\"u}hn, \emph{Charge and Energy Transfer Dynamics in Molecular
  Systems, 2nd, Revised and Enlarged Edition, by Volkhard May, Oliver K{\"u}hn,
  pp. 490. ISBN 3-527-40396-5. Wiley-VCH, February 2004.}, 2004,  490\relax
\mciteBstWouldAddEndPuncttrue
\mciteSetBstMidEndSepPunct{\mcitedefaultmidpunct}
{\mcitedefaultendpunct}{\mcitedefaultseppunct}\relax
\EndOfBibitem
\bibitem[Michetti and La~Rocca(2005)]{michetti_polariton_2005}
P.~Michetti and G.~C. La~Rocca, \emph{Physical Review B}, 2005, \textbf{71},
  115320\relax
\mciteBstWouldAddEndPuncttrue
\mciteSetBstMidEndSepPunct{\mcitedefaultmidpunct}
{\mcitedefaultendpunct}{\mcitedefaultseppunct}\relax
\EndOfBibitem
\bibitem[Michetti and La~Rocca(2008)]{michetti_polariton_2008}
P.~Michetti and G.~C. La~Rocca, \emph{Physica E: Low-dimensional Systems and
  Nanostructures}, 2008, \textbf{40}, 1926--1929\relax
\mciteBstWouldAddEndPuncttrue
\mciteSetBstMidEndSepPunct{\mcitedefaultmidpunct}
{\mcitedefaultendpunct}{\mcitedefaultseppunct}\relax
\EndOfBibitem
\bibitem[Holstein(1959)]{holstein1959}
T.~Holstein, \emph{Annals of Physics}, 1959, \textbf{8}, 325--342\relax
\mciteBstWouldAddEndPuncttrue
\mciteSetBstMidEndSepPunct{\mcitedefaultmidpunct}
{\mcitedefaultendpunct}{\mcitedefaultseppunct}\relax
\EndOfBibitem
\bibitem[Herrera and Spano(2017)]{herrera_absorption_2017}
F.~Herrera and F.~C. Spano, \emph{Physical Review A}, 2017, \textbf{95},
  053867\relax
\mciteBstWouldAddEndPuncttrue
\mciteSetBstMidEndSepPunct{\mcitedefaultmidpunct}
{\mcitedefaultendpunct}{\mcitedefaultseppunct}\relax
\EndOfBibitem
\bibitem[{\'C}wik \emph{et~al.}(2016){\'C}wik, Kirton, De~Liberato, and
  Keeling]{cwik_excitonic_2016}
J.~A. {\'C}wik, P.~Kirton, S.~De~Liberato and J.~Keeling, \emph{Physical Review
  A}, 2016, \textbf{93}, 033840\relax
\mciteBstWouldAddEndPuncttrue
\mciteSetBstMidEndSepPunct{\mcitedefaultmidpunct}
{\mcitedefaultendpunct}{\mcitedefaultseppunct}\relax
\EndOfBibitem
\bibitem[Spano(2015)]{spano_optical_2015}
F.~C. Spano, \emph{The Journal of Chemical Physics}, 2015, \textbf{142},
  184707\relax
\mciteBstWouldAddEndPuncttrue
\mciteSetBstMidEndSepPunct{\mcitedefaultmidpunct}
{\mcitedefaultendpunct}{\mcitedefaultseppunct}\relax
\EndOfBibitem
\bibitem[Knapp(1984)]{knapp1984}
E.~W. Knapp, \emph{Chemical Physics}, 1984, \textbf{85}, 73--82\relax
\mciteBstWouldAddEndPuncttrue
\mciteSetBstMidEndSepPunct{\mcitedefaultmidpunct}
{\mcitedefaultendpunct}{\mcitedefaultseppunct}\relax
\EndOfBibitem
\bibitem[Strashko and Keeling(2016)]{strashko2016}
A.~Strashko and J.~Keeling, \emph{Physical Review A}, 2016, \textbf{94},
  023843\relax
\mciteBstWouldAddEndPuncttrue
\mciteSetBstMidEndSepPunct{\mcitedefaultmidpunct}
{\mcitedefaultendpunct}{\mcitedefaultseppunct}\relax
\EndOfBibitem
\bibitem[Nitzan(2006)]{nitzan2006chemical}
A.~Nitzan, \emph{Chemical Dynamics in Condensed Phases: Relaxation, Transfer
  and Reactions in Condensed Molecular Systems}, {Oxford university press},
  2006\relax
\mciteBstWouldAddEndPuncttrue
\mciteSetBstMidEndSepPunct{\mcitedefaultmidpunct}
{\mcitedefaultendpunct}{\mcitedefaultseppunct}\relax
\EndOfBibitem
\bibitem[Gardiner and Haken(1991)]{gardiner1991quantum}
C.~W. Gardiner and H.~Haken, \emph{Quantum Noise}, {Springer Berlin}, 1991,
  vol.~26\relax
\mciteBstWouldAddEndPuncttrue
\mciteSetBstMidEndSepPunct{\mcitedefaultmidpunct}
{\mcitedefaultendpunct}{\mcitedefaultseppunct}\relax
\EndOfBibitem
\bibitem[Mart{\'\i}nez-Mart{\'\i}nez and
  Yuen-Zhou(2018)]{martinez-martinez2018a}
L.~A. Mart{\'\i}nez-Mart{\'\i}nez and J.~Yuen-Zhou, \emph{New Journal of
  Physics}, 2018, \textbf{20}, 018002\relax
\mciteBstWouldAddEndPuncttrue
\mciteSetBstMidEndSepPunct{\mcitedefaultmidpunct}
{\mcitedefaultendpunct}{\mcitedefaultseppunct}\relax
\EndOfBibitem
\bibitem[Savona \emph{et~al.}(1995)Savona, Andreani, Schwendimann, and
  Quattropani]{savona1995}
V.~Savona, L.~C. Andreani, P.~Schwendimann and A.~Quattropani, \emph{Solid
  State Communications}, 1995, \textbf{93}, 733--739\relax
\mciteBstWouldAddEndPuncttrue
\mciteSetBstMidEndSepPunct{\mcitedefaultmidpunct}
{\mcitedefaultendpunct}{\mcitedefaultseppunct}\relax
\EndOfBibitem
\bibitem[Dirac(1927)]{dirac1927}
P.~a.~M. Dirac, \emph{Proc. R. Soc. Lond. A}, 1927, \textbf{114},
  243--265\relax
\mciteBstWouldAddEndPuncttrue
\mciteSetBstMidEndSepPunct{\mcitedefaultmidpunct}
{\mcitedefaultendpunct}{\mcitedefaultseppunct}\relax
\EndOfBibitem
\bibitem[Tartakovskii \emph{et~al.}(2001)Tartakovskii, Emam-Ismail, Lidzey,
  Skolnick, Bradley, Walker, and Agranovich]{tartakovskii2001}
A.~I. Tartakovskii, M.~Emam-Ismail, D.~G. Lidzey, M.~S. Skolnick, D.~D.~C.
  Bradley, S.~Walker and V.~M. Agranovich, \emph{Physical Review B}, 2001,
  \textbf{63}, 121302\relax
\mciteBstWouldAddEndPuncttrue
\mciteSetBstMidEndSepPunct{\mcitedefaultmidpunct}
{\mcitedefaultendpunct}{\mcitedefaultseppunct}\relax
\EndOfBibitem
\bibitem[Agranovich \emph{et~al.}(2002)Agranovich, Litinskaia, and
  Lidzey]{agranovich2002}
V.~Agranovich, M.~Litinskaia and D.~Lidzey, \emph{physica status solidi (b)},
  2002, \textbf{234}, 130--138\relax
\mciteBstWouldAddEndPuncttrue
\mciteSetBstMidEndSepPunct{\mcitedefaultmidpunct}
{\mcitedefaultendpunct}{\mcitedefaultseppunct}\relax
\EndOfBibitem
\bibitem[Michetti and La~Rocca(2009)]{michetti2009}
P.~Michetti and G.~C. La~Rocca, \emph{Physical Review B}, 2009, \textbf{79},
  035325\relax
\mciteBstWouldAddEndPuncttrue
\mciteSetBstMidEndSepPunct{\mcitedefaultmidpunct}
{\mcitedefaultendpunct}{\mcitedefaultseppunct}\relax
\EndOfBibitem
\bibitem[F{\"o}rster(1948)]{forster_zwischenmolekulare_1948}
T.~F{\"o}rster, \emph{Annalen der Physik}, 1948, \textbf{437}, 55--75\relax
\mciteBstWouldAddEndPuncttrue
\mciteSetBstMidEndSepPunct{\mcitedefaultmidpunct}
{\mcitedefaultendpunct}{\mcitedefaultseppunct}\relax
\EndOfBibitem
\bibitem[Medintz and Hildebrandt(2013)]{medintz2013fret}
I.~L. Medintz and N.~Hildebrandt, \emph{{{FRET}}-{{F{\"o}rster}} Resonance
  Energy Transfer: From Theory to Applications}, {John Wiley \& Sons},
  2013\relax
\mciteBstWouldAddEndPuncttrue
\mciteSetBstMidEndSepPunct{\mcitedefaultmidpunct}
{\mcitedefaultendpunct}{\mcitedefaultseppunct}\relax
\EndOfBibitem
\bibitem[Basko \emph{et~al.}(2000)Basko, Bassani, La~Rocca, and
  Agranovich]{basko2000}
D.~M. Basko, F.~Bassani, G.~C. La~Rocca and V.~M. Agranovich, \emph{Physical
  Review B}, 2000, \textbf{62}, 15962--15977\relax
\mciteBstWouldAddEndPuncttrue
\mciteSetBstMidEndSepPunct{\mcitedefaultmidpunct}
{\mcitedefaultendpunct}{\mcitedefaultseppunct}\relax
\EndOfBibitem
\bibitem[Liu \emph{et~al.}(2007)Liu, Long, Choi, Kang, and Lee]{liu2007}
G.~L. Liu, Y.-T. Long, Y.~Choi, T.~Kang and L.~P. Lee, \emph{Nature Methods},
  2007, \textbf{4}, 1015--1017\relax
\mciteBstWouldAddEndPuncttrue
\mciteSetBstMidEndSepPunct{\mcitedefaultmidpunct}
{\mcitedefaultendpunct}{\mcitedefaultseppunct}\relax
\EndOfBibitem
\bibitem[Feist and Garcia-Vidal(2015)]{feist_extraordinary_2015}
J.~Feist and F.~J. Garcia-Vidal, \emph{Physical Review Letters}, 2015,
  \textbf{114}, 196402\relax
\mciteBstWouldAddEndPuncttrue
\mciteSetBstMidEndSepPunct{\mcitedefaultmidpunct}
{\mcitedefaultendpunct}{\mcitedefaultseppunct}\relax
\EndOfBibitem
\bibitem[Schachenmayer \emph{et~al.}(2015)Schachenmayer, Genes, Tignone, and
  Pupillo]{schachenmayer_cavity-enhanced_2015}
J.~Schachenmayer, C.~Genes, E.~Tignone and G.~Pupillo, \emph{Physical Review
  Letters}, 2015, \textbf{114}, 196403\relax
\mciteBstWouldAddEndPuncttrue
\mciteSetBstMidEndSepPunct{\mcitedefaultmidpunct}
{\mcitedefaultendpunct}{\mcitedefaultseppunct}\relax
\EndOfBibitem
\bibitem[Yuen-Zhou \emph{et~al.}(2014)Yuen-Zhou, Saikin, Yao, and
  Aspuru-Guzik]{yuen-zhou2014}
J.~Yuen-Zhou, S.~K. Saikin, N.~Y. Yao and A.~Aspuru-Guzik, \emph{Nature
  Materials}, 2014, \textbf{13}, 1026--1032\relax
\mciteBstWouldAddEndPuncttrue
\mciteSetBstMidEndSepPunct{\mcitedefaultmidpunct}
{\mcitedefaultendpunct}{\mcitedefaultseppunct}\relax
\EndOfBibitem
\bibitem[Yuen-Zhou \emph{et~al.}(2016)Yuen-Zhou, Saikin, Zhu, Onbasli, Ross,
  Bulovic, and Baldo]{yuen-zhou_plexciton_2016}
J.~Yuen-Zhou, S.~K. Saikin, T.~Zhu, M.~C. Onbasli, C.~A. Ross, V.~Bulovic and
  M.~A. Baldo, \emph{Nature Communications}, 2016, \textbf{7}, 11783\relax
\mciteBstWouldAddEndPuncttrue
\mciteSetBstMidEndSepPunct{\mcitedefaultmidpunct}
{\mcitedefaultendpunct}{\mcitedefaultseppunct}\relax
\EndOfBibitem
\bibitem[Smith and Michl(2010)]{smith2010}
M.~B. Smith and J.~Michl, \emph{Chemical Reviews}, 2010, \textbf{110},
  6891--6936\relax
\mciteBstWouldAddEndPuncttrue
\mciteSetBstMidEndSepPunct{\mcitedefaultmidpunct}
{\mcitedefaultendpunct}{\mcitedefaultseppunct}\relax
\EndOfBibitem
\bibitem[Smith and Michl(2013)]{smith2013}
M.~B. Smith and J.~Michl, \emph{Annual Review of Physical Chemistry}, 2013,
  \textbf{64}, 361--386\relax
\mciteBstWouldAddEndPuncttrue
\mciteSetBstMidEndSepPunct{\mcitedefaultmidpunct}
{\mcitedefaultendpunct}{\mcitedefaultseppunct}\relax
\EndOfBibitem
\bibitem[Congreve \emph{et~al.}(2013)Congreve, Lee, Thompson, Hontz, Yost,
  Reusswig, Bahlke, Reineke, Voorhis, and Baldo]{congreve2013}
D.~N. Congreve, J.~Lee, N.~J. Thompson, E.~Hontz, S.~R. Yost, P.~D. Reusswig,
  M.~E. Bahlke, S.~Reineke, T.~V. Voorhis and M.~A. Baldo, \emph{Science},
  2013, \textbf{340}, 334--337\relax
\mciteBstWouldAddEndPuncttrue
\mciteSetBstMidEndSepPunct{\mcitedefaultmidpunct}
{\mcitedefaultendpunct}{\mcitedefaultseppunct}\relax
\EndOfBibitem
\bibitem[Yost \emph{et~al.}(2014)Yost, Lee, Wilson, Wu, McMahon, Parkhurst,
  Thompson, Congreve, Rao, Johnson, Sfeir, Bawendi, Swager, Friend, Baldo, and
  Voorhis]{yost2014}
S.~R. Yost, J.~Lee, M.~W.~B. Wilson, T.~Wu, D.~P. McMahon, R.~R. Parkhurst,
  N.~J. Thompson, D.~N. Congreve, A.~Rao, K.~Johnson, M.~Y. Sfeir, M.~G.
  Bawendi, T.~M. Swager, R.~H. Friend, M.~A. Baldo and T.~V. Voorhis,
  \emph{Nature Chemistry}, 2014, \textbf{6}, 492\relax
\mciteBstWouldAddEndPuncttrue
\mciteSetBstMidEndSepPunct{\mcitedefaultmidpunct}
{\mcitedefaultendpunct}{\mcitedefaultseppunct}\relax
\EndOfBibitem
\bibitem[Jortner(1976)]{jortner1976}
J.~Jortner, \emph{The Journal of Chemical Physics}, 1976, \textbf{64},
  4860--4867\relax
\mciteBstWouldAddEndPuncttrue
\mciteSetBstMidEndSepPunct{\mcitedefaultmidpunct}
{\mcitedefaultendpunct}{\mcitedefaultseppunct}\relax
\EndOfBibitem
\bibitem[Xiang \emph{et~al.}(2017)Xiang, Ribeiro, Dunkelberger, Wang, Li,
  Simpkins, Owrutsky, Yuen-Zhou, and Xiong]{xiang2017}
B.~Xiang, R.~F. Ribeiro, A.~D. Dunkelberger, J.~Wang, Y.~Li, B.~S. Simpkins,
  J.~C. Owrutsky, J.~Yuen-Zhou and W.~Xiong, \emph{arXiv:1711.11222 [cond-mat,
  physics:physics, physics:quant-ph]}, 2017\relax
\mciteBstWouldAddEndPuncttrue
\mciteSetBstMidEndSepPunct{\mcitedefaultmidpunct}
{\mcitedefaultendpunct}{\mcitedefaultseppunct}\relax
\EndOfBibitem
\bibitem[Henry and Hopfield(1965)]{henry1965}
C.~H. Henry and J.~J. Hopfield, \emph{Physical Review Letters}, 1965,
  \textbf{15}, 964--966\relax
\mciteBstWouldAddEndPuncttrue
\mciteSetBstMidEndSepPunct{\mcitedefaultmidpunct}
{\mcitedefaultendpunct}{\mcitedefaultseppunct}\relax
\EndOfBibitem
\bibitem[Mills and Burstein(1974)]{mills1974}
D.~L. Mills and E.~Burstein, \emph{Reports on Progress in Physics}, 1974,
  \textbf{37}, 817\relax
\mciteBstWouldAddEndPuncttrue
\mciteSetBstMidEndSepPunct{\mcitedefaultmidpunct}
{\mcitedefaultendpunct}{\mcitedefaultseppunct}\relax
\EndOfBibitem
\bibitem[Denisov \emph{et~al.}(1987)Denisov, Mavrin, and
  Podobedov]{denisov1987}
V.~N. Denisov, B.~N. Mavrin and V.~B. Podobedov, \emph{Physics Reports}, 1987,
  \textbf{151}, 1--92\relax
\mciteBstWouldAddEndPuncttrue
\mciteSetBstMidEndSepPunct{\mcitedefaultmidpunct}
{\mcitedefaultendpunct}{\mcitedefaultseppunct}\relax
\EndOfBibitem
\bibitem[Shalabney \emph{et~al.}(2015)Shalabney, George, Hutchison, Pupillo,
  Genet, and Ebbesen]{shalabney_coherent_2015}
A.~Shalabney, J.~George, J.~Hutchison, G.~Pupillo, C.~Genet and T.~W. Ebbesen,
  \emph{Nature Communications}, 2015, \textbf{6}:5981 \relax
\mciteBstWouldAddEndPuncttrue
\mciteSetBstMidEndSepPunct{\mcitedefaultmidpunct}
{\mcitedefaultendpunct}{\mcitedefaultseppunct}\relax
\EndOfBibitem
\bibitem[George \emph{et~al.}(2015)George, Shalabney, Hutchison, Genet, and
  Ebbesen]{george_liquid-phase_2015}
J.~George, A.~Shalabney, J.~A. Hutchison, C.~Genet and T.~W. Ebbesen, \emph{The
  Journal of Physical Chemistry Letters}, 2015, \textbf{6}, 1027--1031\relax
\mciteBstWouldAddEndPuncttrue
\mciteSetBstMidEndSepPunct{\mcitedefaultmidpunct}
{\mcitedefaultendpunct}{\mcitedefaultseppunct}\relax
\EndOfBibitem
\bibitem[Long and Simpkins(2015)]{long_coherent_2015}
J.~P. Long and B.~S. Simpkins, \emph{ACS Photonics}, 2015, \textbf{2},
  130--136\relax
\mciteBstWouldAddEndPuncttrue
\mciteSetBstMidEndSepPunct{\mcitedefaultmidpunct}
{\mcitedefaultendpunct}{\mcitedefaultseppunct}\relax
\EndOfBibitem
\bibitem[Muallem \emph{et~al.}(2016)Muallem, Palatnik, Nessim, and
  Tischler]{muallem2016a}
M.~Muallem, A.~Palatnik, G.~D. Nessim and Y.~R. Tischler, \emph{Annalen der
  Physik}, 2016, \textbf{528}, 313--320\relax
\mciteBstWouldAddEndPuncttrue
\mciteSetBstMidEndSepPunct{\mcitedefaultmidpunct}
{\mcitedefaultendpunct}{\mcitedefaultseppunct}\relax
\EndOfBibitem
\bibitem[Muallem \emph{et~al.}(2016)Muallem, Palatnik, Nessim, and
  Tischler]{muallem2016}
M.~Muallem, A.~Palatnik, G.~D. Nessim and Y.~R. Tischler, \emph{The Journal of
  Physical Chemistry Letters}, 2016, \textbf{7}, 2002--2008\relax
\mciteBstWouldAddEndPuncttrue
\mciteSetBstMidEndSepPunct{\mcitedefaultmidpunct}
{\mcitedefaultendpunct}{\mcitedefaultseppunct}\relax
\EndOfBibitem
\bibitem[Casey and Sparks(2016)]{casey_vibrational_2016}
S.~R. Casey and J.~R. Sparks, \emph{The Journal of Physical Chemistry C}, 2016,
  \textbf{120}, 28138--28143\relax
\mciteBstWouldAddEndPuncttrue
\mciteSetBstMidEndSepPunct{\mcitedefaultmidpunct}
{\mcitedefaultendpunct}{\mcitedefaultseppunct}\relax
\EndOfBibitem
\bibitem[Vergauwe \emph{et~al.}(2016)Vergauwe, George, Chervy, Hutchison,
  Shalabney, Torbeev, and Ebbesen]{vergauwe_quantum_2016}
R.~M.~A. Vergauwe, J.~George, T.~Chervy, J.~A. Hutchison, A.~Shalabney, V.~Y.
  Torbeev and T.~W. Ebbesen, \emph{The Journal of Physical Chemistry Letters},
  2016, \textbf{7}, 4159--4164\relax
\mciteBstWouldAddEndPuncttrue
\mciteSetBstMidEndSepPunct{\mcitedefaultmidpunct}
{\mcitedefaultendpunct}{\mcitedefaultseppunct}\relax
\EndOfBibitem
\bibitem[Dunkelberger \emph{et~al.}(2016)Dunkelberger, Spann, Fears, Simpkins,
  and Owrutsky]{dunkelberger_modified_2016}
A.~D. Dunkelberger, B.~T. Spann, K.~P. Fears, B.~S. Simpkins and J.~C.
  Owrutsky, \emph{Nature Communications}, 2016, \textbf{7}:13504\relax
\mciteBstWouldAddEndPuncttrue
\mciteSetBstMidEndSepPunct{\mcitedefaultmidpunct}
{\mcitedefaultendpunct}{\mcitedefaultseppunct}\relax
\EndOfBibitem
\bibitem[Ahn \emph{et~al.}(2018)Ahn, Vurgaftman, Dunkelberger, Owrutsky, and
  Simpkins]{ahn2018}
W.~Ahn, I.~Vurgaftman, A.~D. Dunkelberger, J.~C. Owrutsky and B.~S. Simpkins,
  \emph{ACS Photonics}, 2018, \textbf{5}, 158--166\relax
\mciteBstWouldAddEndPuncttrue
\mciteSetBstMidEndSepPunct{\mcitedefaultmidpunct}
{\mcitedefaultendpunct}{\mcitedefaultseppunct}\relax
\EndOfBibitem
\bibitem[Kapon \emph{et~al.}(2017)Kapon, Yitzhari, Palatnik, and
  Tischler]{kapon_vibrational_2017}
O.~Kapon, R.~Yitzhari, A.~Palatnik and Y.~R. Tischler, \emph{The Journal of
  Physical Chemistry C}, 2017, \textbf{121}, 18845--18853\relax
\mciteBstWouldAddEndPuncttrue
\mciteSetBstMidEndSepPunct{\mcitedefaultmidpunct}
{\mcitedefaultendpunct}{\mcitedefaultseppunct}\relax
\EndOfBibitem
\bibitem[Memmi \emph{et~al.}(2017)Memmi, Benson, Sadofev, and
  Kalusniak]{memmi2017}
H.~Memmi, O.~Benson, S.~Sadofev and S.~Kalusniak, \emph{Physical Review
  Letters}, 2017, \textbf{118}, 126802\relax
\mciteBstWouldAddEndPuncttrue
\mciteSetBstMidEndSepPunct{\mcitedefaultmidpunct}
{\mcitedefaultendpunct}{\mcitedefaultseppunct}\relax
\EndOfBibitem
\bibitem[Fano(1956)]{fano1956}
U.~Fano, \emph{Physical Review}, 1956, \textbf{103}, 1202--1218\relax
\mciteBstWouldAddEndPuncttrue
\mciteSetBstMidEndSepPunct{\mcitedefaultmidpunct}
{\mcitedefaultendpunct}{\mcitedefaultseppunct}\relax
\EndOfBibitem
\bibitem[Mukamel(1999)]{mukamel1999principles}
S.~Mukamel, \emph{Principles of Nonlinear Optical Spectroscopy}, {Oxford
  University Press on Demand}, 1999\relax
\mciteBstWouldAddEndPuncttrue
\mciteSetBstMidEndSepPunct{\mcitedefaultmidpunct}
{\mcitedefaultendpunct}{\mcitedefaultseppunct}\relax
\EndOfBibitem
\bibitem[Hamm and Zanni(2011)]{hamm2011concepts}
P.~Hamm and M.~Zanni, \emph{Concepts and Methods of {{2D}} Infrared
  Spectroscopy}, {Cambridge University Press}, 2011\relax
\mciteBstWouldAddEndPuncttrue
\mciteSetBstMidEndSepPunct{\mcitedefaultmidpunct}
{\mcitedefaultendpunct}{\mcitedefaultseppunct}\relax
\EndOfBibitem
\bibitem[Chervy \emph{et~al.}(2018)Chervy, Thomas, Akiki, Vergauwe, Shalabney,
  George, Devaux, Hutchison, Genet, and Ebbesen]{chervy2018}
T.~Chervy, A.~Thomas, E.~Akiki, R.~M.~A. Vergauwe, A.~Shalabney, J.~George,
  E.~Devaux, J.~A. Hutchison, C.~Genet and T.~W. Ebbesen, \emph{ACS Photonics},
  2018, \textbf{5}, 217--224\relax
\mciteBstWouldAddEndPuncttrue
\mciteSetBstMidEndSepPunct{\mcitedefaultmidpunct}
{\mcitedefaultendpunct}{\mcitedefaultseppunct}\relax
\EndOfBibitem
\bibitem[Herzberg and Spinks(1939)]{herzberg1939molecular}
G.~Herzberg and J.~Spinks, \emph{Molecular {{Spectra}} and {{Molecular
  Structure}}: {{Infrared}} and {{Raman}} Spectra of Polyatomic Molecules},
  {Van Nostrand}, 1939\relax
\mciteBstWouldAddEndPuncttrue
\mciteSetBstMidEndSepPunct{\mcitedefaultmidpunct}
{\mcitedefaultendpunct}{\mcitedefaultseppunct}\relax
\EndOfBibitem
\bibitem[Yuen-Zhou \emph{et~al.}(2014)Yuen-Zhou, Krich, Aspuru-Guzik, Kassal,
  and Johnson]{yuen2014ultrafast}
J.~Yuen-Zhou, J.~Krich, A.~Aspuru-Guzik, I.~Kassal and A.~Johnson,
  \emph{Ultrafast {{Spectroscopy}}: {{Quantum Information}} and
  {{Wavepackets}}}, {Institute of Physics Publishing}, 2014\relax
\mciteBstWouldAddEndPuncttrue
\mciteSetBstMidEndSepPunct{\mcitedefaultmidpunct}
{\mcitedefaultendpunct}{\mcitedefaultseppunct}\relax
\EndOfBibitem
\bibitem[Valleau \emph{et~al.}(2012)Valleau, Saikin, Yung, and
  Guzik]{valleau2012}
S.~Valleau, S.~K. Saikin, M.-H. Yung and A.~A. Guzik, \emph{The Journal of
  Chemical Physics}, 2012, \textbf{137}, 034109\relax
\mciteBstWouldAddEndPuncttrue
\mciteSetBstMidEndSepPunct{\mcitedefaultmidpunct}
{\mcitedefaultendpunct}{\mcitedefaultseppunct}\relax
\EndOfBibitem
\bibitem[Hertzog \emph{et~al.}(2017)Hertzog, Rudquist, Hutchison, George,
  Ebbesen, and B{\"o}rjesson]{hertzog2017}
M.~Hertzog, P.~Rudquist, J.~A. Hutchison, J.~George, T.~W. Ebbesen and
  K.~B{\"o}rjesson, \emph{Chemistry \textendash{} A European Journal}, 2017,
  \textbf{23}, 18166--18170\relax
\mciteBstWouldAddEndPuncttrue
\mciteSetBstMidEndSepPunct{\mcitedefaultmidpunct}
{\mcitedefaultendpunct}{\mcitedefaultseppunct}\relax
\EndOfBibitem
\bibitem[Herzberg(1968)]{gerhard1968infrared}
G.~Herzberg, \emph{Infrared and {{Raman Spectra}} of {{Polyatomic Molecules}}},
  {Van Nostrand}, 1968\relax
\mciteBstWouldAddEndPuncttrue
\mciteSetBstMidEndSepPunct{\mcitedefaultmidpunct}
{\mcitedefaultendpunct}{\mcitedefaultseppunct}\relax
\EndOfBibitem
\bibitem[Khalil \emph{et~al.}(2003)Khalil, Demird{\"o}ven, and
  Tokmakoff]{khalil_coherent_2003}
M.~Khalil, N.~Demird{\"o}ven and A.~Tokmakoff, \emph{The Journal of Physical
  Chemistry A}, 2003, \textbf{107}, 5258--5279\relax
\mciteBstWouldAddEndPuncttrue
\mciteSetBstMidEndSepPunct{\mcitedefaultmidpunct}
{\mcitedefaultendpunct}{\mcitedefaultseppunct}\relax
\EndOfBibitem
\bibitem[Ishii \emph{et~al.}(2009)Ishii, Takeuchi, and Tahara]{ishii2009}
K.~Ishii, S.~Takeuchi and T.~Tahara, \emph{The Journal of Chemical Physics},
  2009, \textbf{131}, 044512\relax
\mciteBstWouldAddEndPuncttrue
\mciteSetBstMidEndSepPunct{\mcitedefaultmidpunct}
{\mcitedefaultendpunct}{\mcitedefaultseppunct}\relax
\EndOfBibitem
\bibitem[Ciuti \emph{et~al.}(2005)Ciuti, Bastard, and
  Carusotto]{ciuti_quantum_2005}
C.~Ciuti, G.~Bastard and I.~Carusotto, \emph{Physical Review B}, 2005,
  \textbf{72}, 115303\relax
\mciteBstWouldAddEndPuncttrue
\mciteSetBstMidEndSepPunct{\mcitedefaultmidpunct}
{\mcitedefaultendpunct}{\mcitedefaultseppunct}\relax
\EndOfBibitem
\bibitem[Ciuti and Carusotto(2006)]{ciuti_input-output_2006}
C.~Ciuti and I.~Carusotto, \emph{Physical Review A}, 2006, \textbf{74},
  033811\relax
\mciteBstWouldAddEndPuncttrue
\mciteSetBstMidEndSepPunct{\mcitedefaultmidpunct}
{\mcitedefaultendpunct}{\mcitedefaultseppunct}\relax
\EndOfBibitem
\bibitem[Moroz(2014)]{moroz2014}
A.~Moroz, \emph{Annals of Physics}, 2014, \textbf{340}, 252--266\relax
\mciteBstWouldAddEndPuncttrue
\mciteSetBstMidEndSepPunct{\mcitedefaultmidpunct}
{\mcitedefaultendpunct}{\mcitedefaultseppunct}\relax
\EndOfBibitem
\bibitem[K{\'e}na-Cohen \emph{et~al.}(2013)K{\'e}na-Cohen, Maier, and
  Bradley]{kena-cohen2013}
S.~K{\'e}na-Cohen, S.~A. Maier and D.~D.~C. Bradley, \emph{Advanced Optical
  Materials}, 2013, \textbf{1}, 827--833\relax
\mciteBstWouldAddEndPuncttrue
\mciteSetBstMidEndSepPunct{\mcitedefaultmidpunct}
{\mcitedefaultendpunct}{\mcitedefaultseppunct}\relax
\EndOfBibitem
\bibitem[Balci(2013)]{balci2013}
S.~Balci, \emph{Optics Letters}, 2013, \textbf{38}, 4498--4501\relax
\mciteBstWouldAddEndPuncttrue
\mciteSetBstMidEndSepPunct{\mcitedefaultmidpunct}
{\mcitedefaultendpunct}{\mcitedefaultseppunct}\relax
\EndOfBibitem
\bibitem[Gambino \emph{et~al.}(2014)Gambino, Mazzeo, Genco, Di~Stefano,
  Savasta, Patan{\`e}, Ballarini, Mangione, Lerario, Sanvitto, and
  Gigli]{gambino2014}
S.~Gambino, M.~Mazzeo, A.~Genco, O.~Di~Stefano, S.~Savasta, S.~Patan{\`e},
  D.~Ballarini, F.~Mangione, G.~Lerario, D.~Sanvitto and G.~Gigli, \emph{ACS
  Photonics}, 2014, \textbf{1}, 1042--1048\relax
\mciteBstWouldAddEndPuncttrue
\mciteSetBstMidEndSepPunct{\mcitedefaultmidpunct}
{\mcitedefaultendpunct}{\mcitedefaultseppunct}\relax
\EndOfBibitem
\bibitem[George \emph{et~al.}(2015)George, Wang, Chervy, Canaguier-Durand,
  Schaeffer, Lehn, Hutchison, Genet, and Ebbesen]{george_ultra-strong_2015}
J.~George, S.~Wang, T.~Chervy, A.~Canaguier-Durand, G.~Schaeffer, J.-M. Lehn,
  J.~A. Hutchison, C.~Genet and T.~W. Ebbesen, \emph{Faraday Discussions},
  2015, \textbf{178}, 281--294\relax
\mciteBstWouldAddEndPuncttrue
\mciteSetBstMidEndSepPunct{\mcitedefaultmidpunct}
{\mcitedefaultendpunct}{\mcitedefaultseppunct}\relax
\EndOfBibitem
\bibitem[George \emph{et~al.}(2016)George, Chervy, Shalabney, Devaux, Hiura,
  Genet, and Ebbesen]{george_multiple_2016}
J.~George, T.~Chervy, A.~Shalabney, E.~Devaux, H.~Hiura, C.~Genet and T.~W.
  Ebbesen, \emph{Physical Review Letters}, 2016, \textbf{117}, 153601\relax
\mciteBstWouldAddEndPuncttrue
\mciteSetBstMidEndSepPunct{\mcitedefaultmidpunct}
{\mcitedefaultendpunct}{\mcitedefaultseppunct}\relax
\EndOfBibitem
\bibitem[De~Liberato(2014)]{deliberato2014}
S.~De~Liberato, \emph{Physical Review Letters}, 2014, \textbf{112},
  016401\relax
\mciteBstWouldAddEndPuncttrue
\mciteSetBstMidEndSepPunct{\mcitedefaultmidpunct}
{\mcitedefaultendpunct}{\mcitedefaultseppunct}\relax
\EndOfBibitem
\bibitem[Flick \emph{et~al.}(2017)Flick, Ruggenthaler, Appel, and
  Rubio]{flick2017}
J.~Flick, M.~Ruggenthaler, H.~Appel and A.~Rubio, \emph{Proceedings of the
  National Academy of Sciences}, 2017, \textbf{114}, 3026--3034\relax
\mciteBstWouldAddEndPuncttrue
\mciteSetBstMidEndSepPunct{\mcitedefaultmidpunct}
{\mcitedefaultendpunct}{\mcitedefaultseppunct}\relax
\EndOfBibitem
\bibitem[Luk \emph{et~al.}(2017)Luk, Feist, Toppari, and
  Groenhof]{luk_multiscale_2017}
H.~L. Luk, J.~Feist, J.~J. Toppari and G.~Groenhof, \emph{Journal of Chemical
  Theory and Computation}, 2017, \textbf{13}, 4324--4335\relax
\mciteBstWouldAddEndPuncttrue
\mciteSetBstMidEndSepPunct{\mcitedefaultmidpunct}
{\mcitedefaultendpunct}{\mcitedefaultseppunct}\relax
\EndOfBibitem
\bibitem[Vendrell(2017)]{vendrell2017coherent}
O.~Vendrell, \emph{arXiv preprint arXiv:1712.03466}, 2017\relax
\mciteBstWouldAddEndPuncttrue
\mciteSetBstMidEndSepPunct{\mcitedefaultmidpunct}
{\mcitedefaultendpunct}{\mcitedefaultseppunct}\relax
\EndOfBibitem
\bibitem[Subotnik \emph{et~al.}(2016)Subotnik, Jain, Landry, Petit, Ouyang, and
  Bellonzi]{subotnik2016}
J.~E. Subotnik, A.~Jain, B.~Landry, A.~Petit, W.~Ouyang and N.~Bellonzi,
  \emph{Annual Review of Physical Chemistry}, 2016, \textbf{67}, 387--417\relax
\mciteBstWouldAddEndPuncttrue
\mciteSetBstMidEndSepPunct{\mcitedefaultmidpunct}
{\mcitedefaultendpunct}{\mcitedefaultseppunct}\relax
\EndOfBibitem
\bibitem[Makri(1995)]{makri1995}
N.~Makri, \emph{Journal of Mathematical Physics}, 1995, \textbf{36},
  2430--2457\relax
\mciteBstWouldAddEndPuncttrue
\mciteSetBstMidEndSepPunct{\mcitedefaultmidpunct}
{\mcitedefaultendpunct}{\mcitedefaultseppunct}\relax
\EndOfBibitem
\bibitem[Tanimura(2006)]{tanimura2006}
Y.~Tanimura, \emph{Journal of the Physical Society of Japan}, 2006,
  \textbf{75}, 082001\relax
\mciteBstWouldAddEndPuncttrue
\mciteSetBstMidEndSepPunct{\mcitedefaultmidpunct}
{\mcitedefaultendpunct}{\mcitedefaultseppunct}\relax
\EndOfBibitem
\end{mcitethebibliography}
\bibliographystyle{rsc} 

\end{document}